\documentclass[11pt]{article}
\usepackage{amsmath,amssymb,color,graphics,epsfig}
\usepackage{graphicx}
\usepackage{subfigure}

\usepackage{amsfonts}

\newcommand{\be}{\begin{equation}}
\newcommand{\ee}{\end{equation}}
\newcommand{\bea}{\setlength\arraycolsep{2pt} \begin{eqnarray}}
\newcommand{\eea}{\end{eqnarray}}
\newcommand{\nn}{\nonumber}
\newcommand{\mm}{\mathrm}
\newcommand{\mc}{\mathcal}

\def\ft#1#2{{\textstyle{\frac{\scriptstyle #1}{\scriptstyle #2} } }}
\def\fft#1#2{{\frac{#1}{#2}}}

\def\0{{\sst{(0)}}}
\def\1{{\sst{(1)}}}
\def\2{{\sst{(2)}}}
\def\3{{\sst{(3)}}}
\def\4{{\sst{(4)}}}
\def\5{{\sst{(5)}}}
\def\6{{\sst{(6)}}}
\def\7{{\sst{(7)}}}
\def\8{{\sst{(8)}}}
\def\sst#1{{\scriptscriptstyle #1}}

\begin{document}

\begin{flushright}
\end{flushright}

\vspace{25pt}
\begin{center}
{\large {\bf Bulk entanglement and its shape dependence }}

\vspace{10pt}
 Zhong-Ying Fan$^1$\\

\vspace{10pt}
$^1${ Department of Astrophysics, School of Physics and Materials Science, \\
 Guangzhou University, Guangzhou 510006, China }\\
\smallskip

\vspace{40pt}

\underline{ABSTRACT}
\end{center}
We study one-loop bulk entanglement entropy in even spacetime dimensions using the heat kernel method, which captures the universal piece of entanglement entropy, a logarithmically divergent term in even dimensions. In four dimensions, we perform explicit calculations for various shapes of boundary subregions. In particular, for a cusp subregion with an arbitrary opening angle, we find that the bulk entanglement entropy always encodes the same universal information about the boundary theories as the leading entanglement entropy in the large N limit, up to a fixed proportional constant. By smoothly deforming a circle in the boundary, we find that to leading order of the deformations, the bulk entanglement entropy shares the same shape dependence as the leading entanglement entropy and hence the same physical information can be extracted from both cases. This establishes an interesting local/nonlocal duality for holographic $\mm{CFT}_3$. However, the result does not hold for higher dimensional holographic theories.

\vfill {\footnotesize  Email: fanzhy@gzhu.edu.cn\,.}

\thispagestyle{empty}

\pagebreak

\tableofcontents
\addtocontents{toc}{\protect\setcounter{tocdepth}{2}}




\section{Introduction}

Entanglement entropy measures how closely entangled a given wave function is in a quantum mechanical system. It plays an important role in our exploration for a better understanding of quantum systems. In the past two decades, holographic description of entanglement entropy in gauge/gravity duality attracted a lot of attentions since the pioneer work \cite{Ryu:2006bv,Hubeny:2007xt}. The exciting development in this area gives people confidence that this quantity may provide a bridge to connect several different research areas: quantum gravities, quantum field theories and quantum information theories. In particular, recent progress in black hole evaporation \cite{Penington:2019npb,Almheiri:2019psf,Almheiri:2019hni} shows that entanglement entropy might be the correct quantity to characterize evaporation of black holes since it follows the Page curve and hence resolves the information loss paradox argued by Hawking
\cite{Hawking:1974rv,Hawking:1974sw}.

It was first proposed in \cite{Ryu:2006bv,Hubeny:2007xt} that for a boundary subregion A, holographic entanglement entropy is given by one quarter of the area of the minimal area surface $\Sigma$, which is anchored on the boundary and is homologous to A ($\partial \Sigma=\partial A$)
\be S=\fft{\mm{Area}(\Sigma)}{4G_N}  \,.\ee
Without confusion, we always omit the subscript A for $\Sigma$ and $S$ since we will only study the entanglement entropy for a single subregion in the boundary. This is referred to as RT formula in the literature. It looks very similar to that of Bekenstein-Hawking entropy of black holes. For the latter case, the entropy formula can be discovered by introducing a thermodynamic interpretation for Euclidean gravity solutions, which have a $U(1)$ isometry \cite{Gibbons:1976ue}. The partition function for the gravity systems can be evaluated as a Euclidean functional
\be Z(\beta)=\int [\mc{D}\varphi][\mc{D}g_{\mu\nu}]\,\mm{exp}\Big[-I_{grav}-I_{matt} \Big] \,,\ee
where $I_{grav}$ and $I_{matt}$ denotes the Euclidean action of gravity sector and matter sector, respectively. In the saddle point approximation,
 the Euclidean gravitational action can be considered as
 \be I=I_{grav}+I_{matt}=-\log {Z(\beta)}\,,\ee
 and the entropy is derived as
\bea
S=-\big(\beta\partial_\beta -1\big)\log {Z(\beta)}\Big|_{\beta=\beta_H} \,,
\eea
where $\beta_H=1/2\pi T$, $T$ is the temperature of black holes. Application of the above relations to stationary solutions with a $U(1)$ isometry indeed leads to the entropy formula.

In \cite{Lewkowycz:2013nqa}, the approach was successfully generalised to gravity solutions, which in general do not have a $U(1)$ isometry. The basic idea is to perform replica trick in the boundary and extend it to the bulk: the original gravity solution $\mc{M}$ with the boundary subregion $A$ is replicated to $\mc{M}_n$, which is defined by taking $n$ copies of the original manifold, cutting them apart at $A$ and gluing them together in a cyclic order. The entanglement entropy is evaluated as
\be\label{gravityentropy} S=\big(n\partial_n-1 \big)I[\mc{M}_n]\Big|_{n\rightarrow 1}=\partial_n I[\hat{\mc{M}}_n]\Big|_{n\rightarrow 1}\,,\ee
where $\hat{\mc{M}}_n$ stands for the orbifold geometry $\mc{M}_n/Z_n$ and the second equality follows from the relation $I[\mc{M}_n]=n\, I[\hat{\mc{M}}_n]$ because of $Z_n$ symmetry of the replica geometry $\mc{M}_n$. Then a careful examination of the Euclidean action in the orbifold geometry leads to the RT formula \cite{Lewkowycz:2013nqa}. Derivation of holographic entanglement entropy for higher derivative gravities was studied by several different authors using the same approach \cite{Chen:2013qma,Bhattacharyya:2013jma,Dong:2013qoa,Camps:2013zua,Bhattacharyya:2014yga,Miao:2014nxa}. Proof of the formula for some special cases can be found in \cite{Casini:2011kv,Hartman:2013mia,Faulkner:2013yia}. The covariant version of the formula \cite{Hubeny:2007xt} was proved in \cite{Dong:2016hjy}.

In this paper, we are interested in studying bulk one-loop corrections to holographic entanglement entropy (we do not consider backreaction effects from bulk quantum fluctuations, which changes holographic entanglement entropy at order unity as well). This topic was early discussed in \cite{Faulkner:2013ana,Engelhardt:2014gca} and was analyzed extensively in $\mm{AdS}_3/\mm{CFT}_2$ correspondence \cite{Barrella:2013wja,Belin:2019mlt,Agon:2020fqs,Chen:2019xpb}. Quantum corrections to holographic mutual information was studied in \cite{Agon:2015ftl,Chen:2017yns}.

In this paper, we will study bulk entanglement entropy in higher $D\geq 4$ dimensions by using heat kernel method. The method is a powerful tool to capture short distance divergences of one loop effective action in a fixed background. It is particularly useful for even spacetime dimensions, in which case the universal piece of entanglement entropy is a logarithmically divergent term, see for example \cite{Ryu:2006ef,Nishioka:2009un}.  However, in odd dimensions, the universal piece of entanglement entropy is a constant and hence cannot be extracted using the heat kernel method (in this case, bulk entanglement entropy for free scalar fields across hemispheres was studied in \cite{Sugishita:2016iel} using a different approach).

We perform explicit calculations in the $D=4$ dimension for several different shapes of boundary subregions. In particular, for a cusp subregion with an arbitrary opening angle, we find that the bulk entanglement entropy always encodes the same universal information about the boundary theories as the leading entanglement entropy in the large N limit, up to a fixed constant proportional to the central charges of the boundary. Furthermore, by studying a smoothly deformed circle in the boundary, we find that the bulk entanglement entropy, which is non-geometric, shares the same shape dependence with the leading, geometric RT formula. The former captures local information about the $O(1)$ degrees of freedoms (d.o.fs) whilst the latter encodes nonlocal information about the $O(N^2)$ d.o.fs in the boundary. Hence, our result establishes a local/nonlocal duality for the shape dependence of entanglement entropy for holographic $\mm{CFT}_3$. We extend our discussions to the $D=6$ dimension. However, we find that the result in general does not hold any longer for higher dimensions.

The remaining of this paper is organized as follows. In section 2, we apply the heat kernel method to massless scalar fields and discuss the one-loop bulk entanglement entropy in diverse dimensions. We also briefly review the derivations for integrals of curvature invariants in the orbifold geometry. In section 3, we perform the one-loop calculations explicitly in the $D=4$ dimension. In section 4, we extend the calculations to the $D=6$ dimension. We conclude in section 5.

\section{Heat kernel expansion and bulk entanglement }

Extending (\ref{gravityentropy}) to include the contribution of quantum fluctuations in a fixed background, the one-loop bulk entanglement entropy is given by
\be S_q=\big(n\partial_n -1\big)I_{eff}[\mc{M}_n]\Big|_{n\rightarrow 1}=\partial_n I_{eff}[\hat{\mc{M}}_n]\Big|_{n\rightarrow 1}  \label{master}\,,\ee
where $I_{eff}[\mc{M}_n]$ stands for the one-loop effective action in the replica geometry for various quantum fluctuations. In this section, we will adopt heat kernel method to evaluate $I_{eff}$ as well as the bulk entanglement entropy. For convenience, we will work in Euclidean signature throughout this paper.

\subsection{Heat kernel coefficients}

Without loss of generality, we consider a bulk massless scalar field in general dimensions
\be I_{matt}=-\fft12\int \mm{d}^Dx\,\sqrt{G}\,\big( \partial\Phi\big)^2 \,.\ee
The one-loop effective action is formally given by
\be I_{eff}[\mc{M}]=\fft12\log{\mm{det}\Box} \,,\quad \Box=\nabla_\mu \nabla^\mu\,.\ee
In DeWitt-Schwinger proper time representation
\be\log{\mm{det}\Box}=-\int_{\epsilon^2}^\infty \fft{ds}{s}\,\mm{Tr}\big( e^{-s\Box} \big)  \,,\ee
where the heat kernel $\mm{Tr}\big( e^{-s\Box} \big)$ can be expanded as
\be\label{heatkernel} \mm{Tr}\big( e^{-s\Box} \big)=\fft{1}{\big(4\pi s \big)^{D/2}}\sum_{\ell=0}^\infty a_\ell s^\ell \,.\ee
The expansion coefficients $a_\ell$'s can be expressed as geometric invariants in the background manifold. We present some lower lying examples
\bea\label{ans}
&& a_0=\int d^Dx\,\sqrt{G}\,,\nn\\
&& a_1=\fft16\int d^Dx\,\sqrt{G}\,R\,,\nn\\
&& a_2=\int d^Dx\,\sqrt{G}\Big(\fft{1}{72}R^2-\fft{1}{180}R_{\mu\nu}R^{\mu\nu}+\fft{1}{180}R_{\mu\nu\alpha\beta}R^{\mu\nu\alpha\beta}-\fft{1}{30}\Box R \Big)\,.
\eea
The higher order coefficients $a_i$ ($i\geq 3$) have much lengthy expressions, see for example \cite{Vassilevich:2003xt} and the references therein. Notice that the $i$-th order coefficient $a_i$ involves integrals of $i$-th order curvature polynomials as well as derivatives of curvatures with the same length dimensions. For other types of massless fields, such as photons and gravitons, the coefficients $a_i$'s are simply changed by constant factors associated to each curvature invariants in the integrals. As a consequence, our discussions are easily generalised to all massless fields. For massive fields, more terms should be included associated to the mass. Yet, generalisation to
this case is straightforward as well and will not change our main results in this paper.

According to the heat kernel expansion (\ref{heatkernel}), one has to relevant orders
\be \mm{Tr}\big( e^{-s\Box} \big)=\fft{1}{\big(4\pi s \big)^{D/2}}\big(a_0+a_1 s+a_2 s^2+\cdots+a_{[D/2]}s^{[D/2]}+\cdots \big) \,,\ee
where at the $[\fft{D}{2}]$-th order, there will be a logarithmically divergent term to the effective action in even spacetime dimensions. In this case, the coefficient $a_{[D/2]}$ is independent of the cut-off and hence contains universal information about the underlying theories. One finds
\be
I_{eff}[\mc{M}]=-\fft{1}{\big(4\pi \big)^{D/2}}\Big[\ft{a_0}{D\epsilon^D}+\ft{a_1}{(D-2)\epsilon^{D-2}}+\cdots+a_{[D/2]}\log{\big(\ft{\Lambda}{\epsilon}\big)}\Big]+\cdots\,,
\ee
where $\Lambda$ is an infrad cut-off and the dots outside the square bracket stands for regular terms. However, in odd dimensions, the $a_{[D/2]}$ term just gives a least divergent term $\sim a_{[D/2]}/\epsilon$ to the effective action. In this case, the physical information is encoded in the constant term, which however cannot be extracted using the above method ( the constant term of bulk entanglement entropy across hemispheres was derived in \cite{Sugishita:2016iel} by using full heat kernel in AdS space ).

To proceed, we need evaluate the heat kernel coefficients in the replica geometry $\mc{M}_n$ (or its orbifold $\hat{\mc{M}}_n=\mc{M}_n/Z_n$) and derive their derivative with respect to the replica parameter $(\partial_n a_\ell)_{n=1}$. This is similar to the derivations of holographic entanglement entropy \cite{Chen:2013qma,Bhattacharyya:2013jma,Dong:2013qoa,Camps:2013zua,Bhattacharyya:2014yga,Miao:2014nxa}, see also \cite{Solodukhin:1994yz,Fursaev:1995ef,Solodukhin:2011gn,Fursaev:2013fta}.
The major result is each of the coefficients $a_\ell$ can be expressed as a regular part $a_\ell^{reg}$ and a singular part $a_{\ell\,,n}$
\be a_\ell=n\big(a_\ell^{reg}+a_{\ell\,,n}\big) \,.\ee
The regular part is expressed as integrals in the smooth region of $\hat{\mc{M}}_n$ and hence $a_\ell^{reg}$ is independent of $n$. These terms will not contribute to the bulk entanglement entropy $S_q$. On the other hand, the singular part $a_{\ell\,,n}$ is evaluated in the cone region of the orbifold geometry and depends on $n$ nontrivially. Its derivative $\partial_n a_{\ell\,,n}$ in the $n\rightarrow 1$ limit is derived as surface integrals evaluated at the minimal surface $\Sigma$ in the original geometry. Explicitly speaking, if we let $a_\ell=\int d^D x\sqrt{G}\,L(G\,,R\,,\nabla R\,,\cdots)$, one has
\bea
&&a_\ell^{reg}= \int_{\mc{M}} d^D x\sqrt{G}\,L(G\,,R\,,\nabla R\,,\cdots)\,,\nn\\
&&a_{\ell\,,n}=\int_{\hat{\mc{M}}_n} d^D x\sqrt{G}\,L(G\,,R\,,\nabla R\,,\cdots)\,.
\eea
Compared to (\ref{master}), one finds (we set $D=d+2$)
\be (n\partial_n-1) a_\ell\Big|_{n=1}=\partial_n a_{\ell\,,n}\Big|_{n=1}=2\pi \int_\Sigma d^dy\,\sqrt{\gamma}\, \mathfrak{a} \,,\ee
where explicit results about the surface density $\mathfrak{a}$ will be presented in the next subsection. Here we would like to point out that the first coefficient $a_0$ does not have a singular part. As a consequence, the leading order contribution to the bulk entanglement entropy is determined by $a_1$
\be S_q=\fft{\lambda}{4} \fft{A_\Sigma}{\epsilon^{D-2}}+\cdots\,. \ee
where $\lambda$ is a constant depending on the cut-off. This is the well-known are law for entanglement entropy. Combined with the RT formula, the above divergence can be absorbed by renormalzing the Newton constant as
\be \fft{1}{G_{ren}}=\fft{1}{G}+\fft{\lambda}{\epsilon^{D-2}}  \,.\ee
Likewise, subleading order divergences associated to $a_k$ with $k<[D/2]$ can be absorbed by higher order coupling constants in the gravity (counter term) action. For the sake of convenience, we will not repeat this step in the remaining of this paper any longer. Instead, we focus on computing the coefficient $a_{[D/2]}$, which contains universal information about the boundary theories. We introduce
\bea\label{sqeven}
&&S_q=\cdots+s_q\,\log{\big(\ft{\Lambda}{\epsilon}\big)}+\cdots\,,\nn\\
&&s_q=-\fft{1}{(4\pi)^{D/2}}\big(\partial_n a_{[D/2]\,,n} \big)_{n=1}\,.
\eea
Without confusion, $s_q$ will be briefly referred to as bulk entanglement through the remaining of this paper. However, it should be emphasized that $s_q$ diverges in the asymptotic AdS boundary since it will be determined as surface integrals of certain geometric quantities evaluated on the minimal area surface. The divergence structure is similar to that of the leading entanglement entropy in the boundary. Since it is interesting to compare the physical information extracted from both cases, we may set
\be \mc{S}=S/c_{eff}\,,\quad c_{eff}=\fft{\ell^d_{AdS}}{4G} \,,\ee
where $c_{eff}$ is referred to as the effective central charge for the boundary theories. One has for smooth entangling surfaces
\bea
&&\mc{S}=\fft{R^{d-1}}{\delta^{d-1}}+\fft{R^{d-3}}{\delta^{d-3}}+\cdots+\fft{R}{\delta}+(-1)^{\fft{d}{2}}\,s^{\mm{univ}}\,,\nn\\
&&s_q=\fft{R^{d-1}}{\delta^{d-1}}+\fft{R^{d-3}}{\delta^{d-3}}+\cdots+\fft{R}{\delta}+s_q^{\mm{univ}}  \,,
\eea
where $R$ denotes a characteristic length scale of the boundary subregion (it should not be confused with the Ricci scalar) and $\delta$ is the UV cutoff at asymptotic AdS boundary. Universal information about the boundary theories is essentially contained in the above constant pieces. However, for singular subregions, as will be shown in sec.\ref{singular}, one has instead
\bea
&&\mc{S}= \fft{R^{d-1}}{\delta^{d-1}}+\fft{R^{d-3}}{\delta^{d-3}}+\cdots+\fft{R}{\delta}-a(\Omega)\log{\Big(\fft{R}{\delta}\Big)}+\mm{cons}\,,\nn\\
&&s_q=\fft{R^{d-1}}{\delta^{d-1}}+\fft{R^{d-3}}{\delta^{d-3}}+\cdots+\fft{R}{\delta}-b(\Omega)\log{\Big(\fft{R}{\delta}\Big)}+\mm{cons}  \,,\eea
where $\Omega$ stands for the opening angle of the cone in the boundary subregion. In this case, the constant terms are regulator dependent. Instead, the physical information is encoded in the functions $a(\Omega)\,,b(\Omega)$.

\subsection{Integrals of curvature invariants in the orbifold geometry}
To calculate $(\partial_n a_\ell)_{n=1}$, let us briefly review the derivation of integrals of curvature invariants in the orbifold geometry $\hat{\mc{M}}_n$. More details can be found in the literature \cite{Chen:2013qma,Bhattacharyya:2013jma,Dong:2013qoa,Camps:2013zua,Bhattacharyya:2014yga,Miao:2014nxa}.

Close to a codimension$-2$ spacelike hypersurface $\Sigma$ (not necessarily minimal), the metric of the orbifold geometry $\hat{\mc{M}}_n$ looks like a product form
$\mc{C}_n\times \Sigma$. In the adapted coordinates, one has
\be ds^2=e^{2\phi}\hat{g}_{ab}dx^adx^b+\big( \gamma_{ij}(y)+2K_{aij}x^a+Q_{abij}x^a x^b+\cdots \big)dy^idy^j+\cdots \,,\ee
where $K_{aij}$ are the extrinsic curvatures of $\Sigma$ and $Q_{abij}\equiv \partial_a K_{bij}$. We will express the two dimensional cone $\mc{C}_n$ in three types of coordinates: the Cartesian coordinates $(x^1\,,x^2)$, the cylindrical coordinates $(\rho\,,\tau)$
and the complex coordinates $(z\,,\bar z)$. One has
\be \hat{g}_{ab}dx^adx^b=(dx^1)^2+(dx^2)^2=d\rho^2+\rho^2d\tau^2=dzd\bar{z}  \,,\ee
where $x^1=\rho\cos\tau\,,x^2=\rho\sin\tau, z=\rho e^{i\tau}\,,\bar{z}=\rho e^{-i\tau}$.
The function $\phi$ is given by $\phi=-\epsilon\log{\rho}=-\ft{\epsilon}{2}\log{(z\bar{z})}$, where $\epsilon=1-1/n$. For the above metric, the Riemann tensor to leading order can be computed as
\bea\label{curvature0}
&& R_{abcd}=e^{2\phi}\hat{r}_{abcd}\,,\quad R_{iabc}=0\,,\quad R_{aijk}=\mc{D}_k K_{aij}-\mc{D}_j K_{aik} \,,\nn\\
&& R_{ijab}=\gamma^{kl}(K_{bik}K_{aj\ell}-K_{aik}K_{bj\ell}) \,,\nn\\
&& R_{iajb}=\Gamma^c_{ab}K_{cij}+\gamma^{k\ell}K_{ajk}K_{bi\ell}-Q_{abij}\,,\nn\\
&& R_{ijk\ell}=\mc{R}_{ijk\ell}+e^{-2\phi}(K_{ai\ell}K^a_{jk}-K_{aik}K^a_{j\ell})\,,
\eea
where $\hat{r}_{abcd}\,,\mc{R}_{ijkl}$ stand for the Riemann tensor associated to $\hat{g}_{ab}$ and $\gamma_{ij}$ respectively. More results about curvatures and their covariant derivatives in the orbifold geometry can be found in the literature \cite{Chen:2013qma,Bhattacharyya:2013jma,Dong:2013qoa,Camps:2013zua,Bhattacharyya:2014yga,Miao:2014nxa}. For self-consistency, we collect some relevant results in our Appendix A.

For simplicity, let us first consider the general higher order Riemannian gravities
\be I=\int d^Dx\,\sqrt{G}\,L(G_{\mu\nu}\,;R_{\mu\nu\rho\sigma}) \,.\ee
This case was studied very carefully in \cite{Dong:2013qoa}. One has
\be\label{dong} \partial_{n}I[\hat{\mc{M}}_n]\Big|_{n=1}=2\pi \int_\Sigma\,\Big[ \fft{\partial L}{\partial R_{z\bar{z}z\bar{z}}}+\sum_\alpha \Big(\fft{\partial^2 L}{\partial R_{izjz} \partial R_{k\bar{z}l \bar{z}}}\Big)_\alpha\fft{8K_{zij} K_{\bar{z}kl} }{q_\alpha+1} \Big]\,, \ee
where $\int_\Sigma=\int d^dy\,\sqrt{\gamma}$ and $q_\alpha$ is a constant, counting the total number of $Q_{zzij}\,,Q_{\bar{z}\bar{z}ij}$ and pairs of $K$ in each term of the derivative $\fft{\partial^2 L}{\partial R_{izjz} \partial R_{k\bar{z}l \bar{z}}}$, which is expanded according to (\ref{curvature0}). Notice that the second term in the square bracket, should be evaluated in the original geometry, namely taking the limit $n\rightarrow 1$ in the final.

The above result can also be transformed into a covariant form.  The first term gives rise to the usual Wald-Iyer entropy \cite{Wald:1993nt,Iyer:1994ys}
\be\label{wald} \mm{Wald}= -2\pi \int_\Sigma \,\fft{\partial L}{\partial R_{\mu\nu\rho\sigma} }\varepsilon_{\mu\nu}\varepsilon_{\rho\sigma}\,.\ee
The second term, referred to as anomaly in \cite{Dong:2013qoa}, can be written as
\bea\label{anomaly}
&&\mm{Anomaly}=2\pi\int_\Sigma \, \Big(\fft{\partial^2 L}{\partial R_{\mu_1\rho_1\nu_1\sigma_1} \partial R_{\mu_2\rho_2\nu_2\sigma_2}}\Big)_\alpha\,\fft{2K_{\lambda_1\rho_1\sigma_1} K_{\lambda_2\rho_2\sigma_2} }{q_\alpha+1}\times\nn\\
&&\qquad\qquad\qquad\Big[ (n_{\mu_1\mu_2}n_{\nu_1\nu_2}-\varepsilon_{\mu_1\mu_2}\varepsilon_{\nu_1\nu_2})n^{\lambda_1\lambda_2}
+(n_{\mu_1\mu_2}\varepsilon_{\nu_1\nu_2}+\varepsilon_{\mu_1\mu_2}n_{\nu_1\nu_2})\varepsilon^{\lambda_1\lambda_2} \Big]\,,\eea
where
\bea
&&K_{\beta\rho\sigma}=n_\beta^{(a)}K_{a\rho\sigma}\,,\nn\\
&&\varepsilon_{\mu\nu}=n_\mu^{(a)}n_\nu^{(b)}\epsilon_{ab}\,,\nn\\
&&n_{\mu\nu}=n_\mu^{(a)}n_\nu^{(b)}g_{ab} \,,
\eea
where $\epsilon_{ab}$ is the usual Levi-Civita tensor.

Application of above results to some simple cases is give by \cite{Dong:2013qoa}
\bea\label{application}
&&1):\quad L=L(R)\quad\Longrightarrow\quad\partial_{n} I[\hat{\mc{M}}_n]\Big|_{n=1}=-4\pi\int_\Sigma \,\fft{\partial L}{\partial R} \,,\nn\\
&&2):\quad L=R_{\mu\nu}^2\quad\Longrightarrow\quad \partial_{n} I[\hat{\mc{M}}_n]\Big|_{n=1}=-4\pi\int_\Sigma \,\big(R^a_{\,\,\,\,a}-\fft12 K_a K^a \big)\,,\\
&&3):\quad L=R_{\mu\nu\rho\sigma}^2\quad\Longrightarrow\quad \partial_{n} I[\hat{\mc{M}}_n]\Big|_{n=1}=-4\pi\int_\Sigma \,2\big(R^{ab}_{\quad ab}-K_{aij} K^{aij} \big)\,,\nn
\eea
where
\bea
R^a_{\,\,\,\,a}=R_{\mu\nu}n^{\mu\nu} \,,\quad R^{ab}_{\quad ab}=\fft12 R_{\mu\lambda\nu\rho}\,\varepsilon^{\mu\lambda}\varepsilon^{\nu\rho}=R_{\mu\lambda\nu\rho}n^{\mu\nu} n^{\lambda\rho}\,.
\eea

Moreover, for the most general gravitational action $I=\int_\mc{M} d^Dx\,\sqrt{g}\,L(g_{\mu\nu}\,,R\,,\nabla R\,,\cdots)$, there will be more singular terms emerging in the covariant derivatives of Riemann curvatures. In this case, one must consider all the terms involving $\partial_z\partial_{\bar z}\phi$ as well as $\partial_z \phi \partial_{\bar z}\phi$. As a matter of fact, the metric expansion around the bulk surface $\Sigma$ should be considered more carefully, including all the relevant subleading order terms\footnote{Here we have ignored the so-called {\it splitting problems} discussed in \cite{Miao:2014nxa} , since it does not effect our main results in this paper. We refer the interested readers to that paper for details.   }
\be\label{metricexpansion} ds^2=e^{2\phi}\Big[ dzd\bar z+e^{2\phi}T(\bar z dz-z d\bar z)^2\Big]+2i e^{2\phi}V_i(\bar z dz-z d\bar z)dy^i+(\gamma_{ij}+Q_{ij})dy^idy^j
\,, \ee
where
\bea
&& T=T_1+T_a x^a+O(x^2)\,,\nn\\
&& V_i=U_i+V_{ai}x^a+O(x^2)\,,\nn\\
&& Q_{ij}=2K_{aij}x^a+Q_{abij}x^ax^b+P_{abcij}x^ax^bx^c+O(x^4)\,.
\eea
It was established in \cite{Miao:2014nxa} that holographic entanglement entropy can be formally evaluated as
\be\label{Hee}\partial_n I[\hat{\mc{M}}_n] \Big|_{n=1}=2\pi \int_\Sigma \Big[\fft{\delta L}{\delta \partial_z\partial_{\bar z}\phi}+\sum_\alpha\fft{1}{\beta_\alpha}\fft{\delta}{\delta\partial_{\bar{z}}\phi}\Big( \fft{\delta L}{\delta\partial_z \phi}\Big)_{\partial_z\partial_{\bar z}\phi=0}\Big]_{\epsilon=0} \,,\ee
where $\beta_\alpha$ is a constant. In the above result, the first term is referred to as the generalised Wald entropy whilst the second term gives the general anomaly terms \cite{Miao:2014nxa}.

As an example, for six derivative gravities $I=\int_\mc{M} d^Dx\,\sqrt{g}\,L(g_{\mu\nu}\,,R\,,\nabla R)$, the generalised Wald entropy was derived explicitly as \cite{Miao:2014nxa}
\be\label{waldbar} \mm{\underline{Wald}}=-2\pi\int_\Sigma \Big[\fft{\delta L}{\delta R_{\mu\nu\rho\sigma}}\varepsilon_{\mu\nu}\varepsilon_{\rho\sigma}
-2\,\fft{\partial L}{\partial \nabla_\alpha R_{\mu\rho\nu\sigma} }\,K_{\beta\rho\sigma}\big(n^\beta_{\,\,\,\,\mu}n_{\alpha\nu}-\varepsilon^\beta_{\,\,\,\,\mu}\varepsilon_{\alpha\nu} \big) \Big]\,,\ee
where we have introduced a underline on the l.h.s to distinguish it from the pure Riemannian case. However, derivation of the anomaly terms is much more involved and should be studied in a case-by-case basis. Nevertheless, one may gain an intuitive idea about it from a simple trick: expanding the action in terms of pairs $\partial_z\phi \partial_{\bar z}\phi$
\be I=\sum_\alpha \int dz d\bar z\, C_\alpha e^{-\beta_\alpha \phi}\partial_z\phi \partial_{\bar z}\phi+\cdots \,,\ee
then according to (\ref{Hee}), the anomaly term is given by
\be \mm{\underline{Anomaly}}=2\pi\int_\Sigma\, \fft{C_\alpha}{\beta_\alpha} \,.\ee
Evaluation of the anomaly terms for several special cases can be found in \cite{Miao:2014nxa}. We will adopt the results therein for six derivative gravities to our $D=6$ dimensional calculations.

\section{Application to $D=4$ dimension}

Now let us calculate the one-loop bulk entanglement entropy in the $D=4$ dimension, where the relevant heat kernel coefficient is $a_2$. Using (\ref{application}), we deduce
\bea
\big( \partial_n a_{2\,,n}\big)_{n=1}&=&-4\pi\int_\Sigma d^2y\,\sqrt{\gamma}\,\Big[ \fft{1}{36}R-\fft{1}{180}R_{\mu\nu}n^{\mu\nu}\nn\\
&&\qquad\qquad\qquad\quad+\fft{1}{90}R_{\mu\lambda\nu\rho}n^{\mu\nu} n^{\lambda\rho} -\fft{1}{90}K_{aij}K^{aij} +\fft{1}{360}K_a K^a \Big] \,.
\eea
Here it is worth emphasizing that the total derivative term $\Box R$ in $a_2$ does not contribute to $\big( \partial_n a_{2\,,n}\big)_{n=1}$, as shown in \cite{Dong:2015zba}. The result can be even more simplified by using Gauss-Codazzi identity
\be R=R_\Sigma+2R_{\mu\nu} n^{\mu\nu} -R_{\mu\lambda\nu\rho}n^{\mu\nu} n^{\lambda\rho}-K_aK^a+K_{aij}K^{aij} \,,\ee
where $R_\Sigma$ is the scalar curvature of the bulk surface $\Sigma$. One has
\be\label{master40}
\big( \partial_n a_{2\,,n}\big)_{n=1}=-4\pi\int_\Sigma d^2y\,\sqrt{\gamma}\,\Big[ \fft{1}{90}R_\Sigma+\fft{1}{60}R+\fft{1}{60}R_{\mu\nu} n^{\mu\nu}-\fft{1}{120}K_a K^a \Big] \,,
\ee
where the last term in the square bracket vanishes for minimal area surfaces. Moreover, for vacuum solutions to Einstein's gravity (including Schwarzschild black holes), curvatures take particularly simple forms
\bea\label{vacuumcurvature}
R_{\mu\nu}=-(D-1)\ell^{-2}_{AdS}\,g_{\mu\nu}\,,
\eea
where $\ell_{AdS}$ is AdS radius. This greatly simplifies our calculations in the $D=4$ dimension. One finds
\be\label{master4}
\big( \partial_n a_{2\,,n}\big)_{n=1}=-4\pi\int_\Sigma d^2y\,\sqrt{\gamma}\,\Big[ \fft{1}{90}R_\Sigma-\fft{3}{10}\ell^{-2}_{AdS} \Big] \,.
\ee
It implies that to compute the one-loop bulk entanglement entropy, we just need evaluate the scalar curvature of the RT surface in four dimension. Of course, the same result can be obtained by constructing the adapted coordinates for the RT surface and extracting the extrinsic curvatures explicitly, see section \ref{sec5} for more details. The simplest way to test this is checking
the Gauss-Codazzi identity, which implies for RT surfaces in AdS vacuum
\be R_\Sigma+K_{aij}K^{aij}=-d(d-1)\ell_{AdS}^{-2}\,.\ee

\subsection{Hemispheres}

Consider a hemisphere in $AdS_D$ vacuum. For later purpose, we keep our discussions as general as possible and will return to the $D=4$ dimension when necessary. The readers should not be confused. In the boundary, the entangling surface is a $(d-1)$-dimensional sphere
\be \mc{A}=\{t_E=0\,,0\leq r\leq R\}\,,\ee
where $R$ denotes the radius of the sphere. Under the boundary spherical coordinates $(r\,,\Omega_{d-1})$, the bulk metric reads
\be ds^2=\ell^2_{AdS}\fft{d\xi^2+dt_E^2+dr^2+r^2 d\Omega_{d-1}^2}{\xi^2}\,,\ee
where $d\Omega_{d-1}^2$ is the metric of a unit sphere $\mathbf{S}_{d-1}$. According to the RT formula, the bulk minimal surface is derived as \cite{Ryu:2006bv}
\be \Sigma:\quad \xi^2+r^2=R^2 \,,\ee
which describes a hemisphere. The induced metric on $\Sigma$ reads
\be ds^2_{\Sigma}=\ell^2_{AdS}\,\xi^{-4}\big( R^2 dr^2+\xi^2 r^2 d\Omega_{d-1}^2 \big) \,.\ee
In fact, this describes a uniform $d$-dimensional hyperbolic space with curvature radius $\ell_{AdS}$. To see this, we introduce a new coordinate $\theta$: $r=R\sin\theta\,,\xi=R\cos\theta$ and the hyperbolic coordinate $u$ as: $\sinh{u}\equiv \tan\theta$. The induced metric becomes
\bea
ds^2_{\Sigma}&=&\ft{\ell^2_{AdS}}{\cos^2\theta}\big(d\theta^2+\sin^2\theta d\Omega_{d-1}^2 \big)\nn\\
&=&\ell^2_{AdS} \big( du^2+\sinh^2{u} d\Omega_{d-1}^2\big)\,.
\eea
It might be a surprise that the hemisphere coincides with the event horizon of a hyperbolic black hole with radius $\ell_{AdS}$. In fact, it has vanishing extrinsic curvatures $K_{aij}=0$ as well. The physical meaning of this was clarified in \cite{Casini:2011kv}: for conformal field theories (not necessarily holographic ones) defined in a spherical ball-shaped region with radius $R $, the vacuum state can be unitarily transformed into a thermal bath in a hyperbolic space with curvature radius $R$. Hence, the entanglement entropy across a spherical entangling surface $\mathbf{S}_{d-1}$ is equal to the thermal entropy on the hyperbolic space $\mathbf{H}_{d-1}$. For holographic CFTs, the latter is given by the entropy of a $(d+2)$-dimensional AdS-hyperbolic black holes with temperature $T_0=1/2\pi R$
\be ds^2=\Big(\fft{\rho^2}{R^2}-1 \Big)dt_E^2+\fft{d\rho^2}{\fft{\rho^2}{R^2}-1}+\rho^2 \big(du^2+\sinh^2{u} d\Omega^2_{d-1}\big) \,,\ee
where $R=\ell_{AdS}$.

The area of the hemisphere is given by
\be A_\Sigma=\Omega_{\mathbf{H}_d}\ell_{AdS}^{d}\,,\label{areasigma}\ee
where $\Omega_{\mathbf{H}_d}$ is the area of a unit hyperbolic plane $\mathbf{H}_d$. The induced scalar curvature turns out to be a constant
\be R_\Sigma=-d(d-1)\ell^{-2}_{AdS} \,.\ee
Return to the $D=4$ dimension, the leading entanglement reads
\be \mc{S}=A_\Sigma/\ell^2_{AdS}=\fft{2\pi R}{\delta}-2\pi \,.\ee
Evaluating (\ref{master4}) yields
\be \big( \partial_n a_{2\,,n}\big)_{n=1}=\fft{58\pi}{45}\mc{S} \,,\ee
which leads to
\bea\label{sphere1}
s_q=-\fft{29\mc{S}}{360\pi}=-\fft{29R}{180\delta}+\fft{29}{180} \,.\eea
This simple example clearly shows that $s_q$ shares the same divergence structure as the leading entanglement entropy in the boundary. Furthermore, the relation (\ref{sphere1}) inspires us that the universal information encoded in the bulk entanglement may have a simple relation to that in the leading entanglement entropy. This will be said more precisely when we study the entanglement entropy for deformed spheres in sec.\ref{shape}.

\subsection{Finite temperature corrections}
Finite temperature states of the boundary is dual to AdS black holes, described by
\be ds^2=\fft{\ell^2_{AdS}}{\xi^2}\Big(f(\xi)dt_E^2+\fft{d\xi^2}{f(\xi)}+dr^2+r^2 d\phi^2 \Big) \,,\quad f(\xi)=1-\big( \fft{\xi}{\xi_h}\big)^3\,,\ee
where $\xi_h$ denotes the location of event horizon and $(r\,,\phi)$ are the boundary polar coordinates. The temperature of the black hole is given by $T=3/4\pi \xi_h$.

In this case, the bulk minimal surface will be deformed away from the hemisphere because of temperature corrections. Nevertheless, it still respects the boundary symmetry and is characterized by a function $\xi=\xi(r)$. The induced metric of $\Sigma$ becomes
\be ds^2_\Sigma=\fft{\ell^2_{AdS}}{\xi^2}\Big(\big(1+\ft{\xi'(r)^2}{f(\xi)}\big)dr^2+r^2 d\phi^2 \Big) \,,\ee
so that the area functional is given by
\be A_\Sigma=2\pi\ell^2_{AdS}\int_0^R dr\,\fft{r}{\xi^2}\sqrt{1+\ft{\xi'(r)^2}{f(\xi)}}\,.\ee
Variation of the functional leads to
\be \xi \xi''+\Big(2-\fft{\xi f_\xi}{2f} \Big)\xi'^2+\Big(1+\fft{\xi'^2}{f} \Big)\fft{\xi\xi'}{r}+2f=0  \,,\ee
where $f_\xi\equiv df/d\xi$. We would like to analytically solve the equation for a large thermal scale $\beta=1/T$ with $\beta/R>>1$ and then extract the leading thermal corrections to the bulk entanglement entropy. By straightforward calculations, we obtain
\be \xi(r)=\sqrt{R^2-r^2}-\fft{8\pi^3R^3}{27\beta^3}\,\fft{(R^2-r^2)(2R^2-r^2)}{R^3}+\cdots \,.\ee
The hemisphere in AdS vacuum is deformed by a leading order correction $\sim R^3/\beta^3$. The area of the minimal surface becomes
\be A_\Sigma=\Omega_{\mathbf{H}_2}\ell_{AdS}^2+ \fft{32\pi^4R^3}{27\beta^3}\,\ell_{AdS}^2+\cdots\,,\ee
which indeed receives an extra contribution proportional to $R^3/\beta^3$. Likewise, the scalar curvature of $\Sigma$ is corrected at the same order
\be R_\Sigma=-\fft{2}{\ell_{AdS}^2}\Big(1-\fft{32\pi^3R^3}{27\beta^3}\fft{\big(R^2-r^2\big)^{3/2}(2R^2-3r^2)}{R^5}+\cdots \Big)\,.\ee
Substituting these results into (\ref{master4}) and (\ref{sqeven}), we arrive at
\be s_q=-\fft{29R}{180\delta}+\Big(\fft{29}{180}-\fft{112\pi^3R^3}{1215\beta^3}\Big)+\cdots \,.\ee
It is also interesting to compare it to the leading entanglement
\be \mc{S}=A_\Sigma/\ell^2_{AdS}=\fft{2\pi R}{\delta}-\Big(2\pi-\fft{32\pi^4R^3}{27\beta^3}\Big)+\cdots \,.\ee
It is intriguing to notice that in both cases, the universal term decreases as the temperature increases. We may expect that the bulk entanglement entropy signals quantum/thermal phase transitions as the leading entanglement entropy \cite{Albash:2012pd,Cai:2012sk,Kuang:2014kha,Ling:2015dma,Ling:2016wyr}. This may deserve further investigations.

\subsection{Strips}

We move to consider a striped subregion, preserving $(d-1)$-dimensional translational invariance
\be \mc{A}=\{ \mm{y}\in \mathbb{R}^{d}| y^1\in (-a\,,a)\,,y^i\in (0\,,L)\,\,\mm{for}\,\, i=2\,,3\,,\cdots\,,d \}  \,.\ee
Holographic entanglement entropy for this case has been widely studied in the literature, see for example \cite{Ryu:2006bv}. In AdS vaccum, the bulk minimal surface is given by $\xi=\xi(y^1)$
\be \xi'(y^1)=\fft{\sqrt{\xi_*^{2d}-\xi^{2d}}}{\xi^d}\,,\quad \xi_*=\fft{a\,\Gamma\Big(\fft{1}{2d}\Big)}{\sqrt{\pi}\,\Gamma\Big(\fft{d+1}{2d}\Big)} \,,\ee
where $\xi_*$ is the turning point of the minimal surface. Solving the above equation gives
\be y^1(\xi)=\fft{\xi^{d+1}}{(d+1)\xi_*^d}\,\,{}_2F_1\Big(\fft12\,,\fft{d+1}{2d}\,,\fft{3d+1}{2d}\,,\fft{\xi^{2d}}{\xi_*^{2d}}\Big)-a \,.\ee
The induced metric on $\Sigma$ is given by
\be\label{inducedstrip} ds^2_\Sigma=\fft{\ell^2_{AdS}}{\xi^2}\Big(\ft{\xi_*^{2d}}{\xi^{2d}}(dy^1)^2+(dy^2)^2+\cdots+(dy^d)^2 \Big) \,.\ee
Evaluating the area of $\Sigma$ yields
\bea
A_\Sigma&=&2\ell_{AdS}^dL^{d-1}\int_\delta^{\xi_*}d\xi\,\fft{\xi_*^d}{\xi^d\sqrt{\xi_*^{2d}-\xi^{2d}}}\nn\\
&=&\fft{2\ell_{AdS}^dL^{d-1}}{d-1}\Big(\fft{1}{\delta^{d-1}}-\big(\ft{a}{\xi_*}\big)^d\fft{1}{a^{d-1}} \Big)\,.
\eea
In the $D=4$ dimension, it gives
\be \mc{S}=\fft{2L}{\delta}-\fft{2L}{a}\Big( \fft{a}{\xi_*} \Big)^2 \,.\ee
On the other hand, the scalar curvature of the induced metric (\ref{inducedstrip}) is given by
\bea
R_\Sigma&=&\ft{2\xi^4}{\ell^2_{AdS}\,\xi_*^4}\big(\xi'^2+\xi \xi'' \big)    \nn\\
&=& -\ft{2}{\ell^2_{AdS}}\Big(1+\big(\ft{\xi}{\xi_*} \big)^4\Big)\,.
\eea
Substituting the result into (\ref{master4}) and (\ref{sqeven}), we deduce
\bea
s_q&=&-\ft{29}{360\pi}\mc{S}-\ft{L}{90\pi a}\big(\ft{a}{\xi_*} \big)^2 \nn\\
&=&-\ft{29L}{360\pi\delta}+\ft{27L}{360\pi a}\big(\ft{a}{\xi_*} \big)^2\,.
\eea
Again $s_q$ has a linear relation to $\mc{S}$ and hence the two share the same divergence structure in this case.

\subsection{Singular shapes}\label{singular}
Next, we consider singular shapes of entangling surfaces. We choose a cusp subregion in the boundary with opening angle $\Omega$, which is specified as
\be \mc{A}=\{ t_E=0\,,r>0\,,|\theta|\leq \Omega/2 \} \,.\ee
We shall introduce a large distance cutoff for the subregion $r_{max}=R$. The leading entanglement entropy for this case was extensively studied in \cite{Hirata:2006jx,Bueno:2015rda,Bueno:2015xda}. Following these papers, we parameterized the bulk surface as $\xi=\xi(r\,,\theta)=r \varphi(\theta)$. The separation of variables is due to scaling symmetries of AdS vacuum (and there is no other scale in the problem). The function $\varphi(\theta)$  satisfies $\dot{\varphi}(0)=0\,,\varphi(\pm \Omega/2)=0$, where $\dot{\varphi}\equiv d\varphi/d\theta$.

\begin{figure}
  \centering
  \includegraphics[width=210pt]{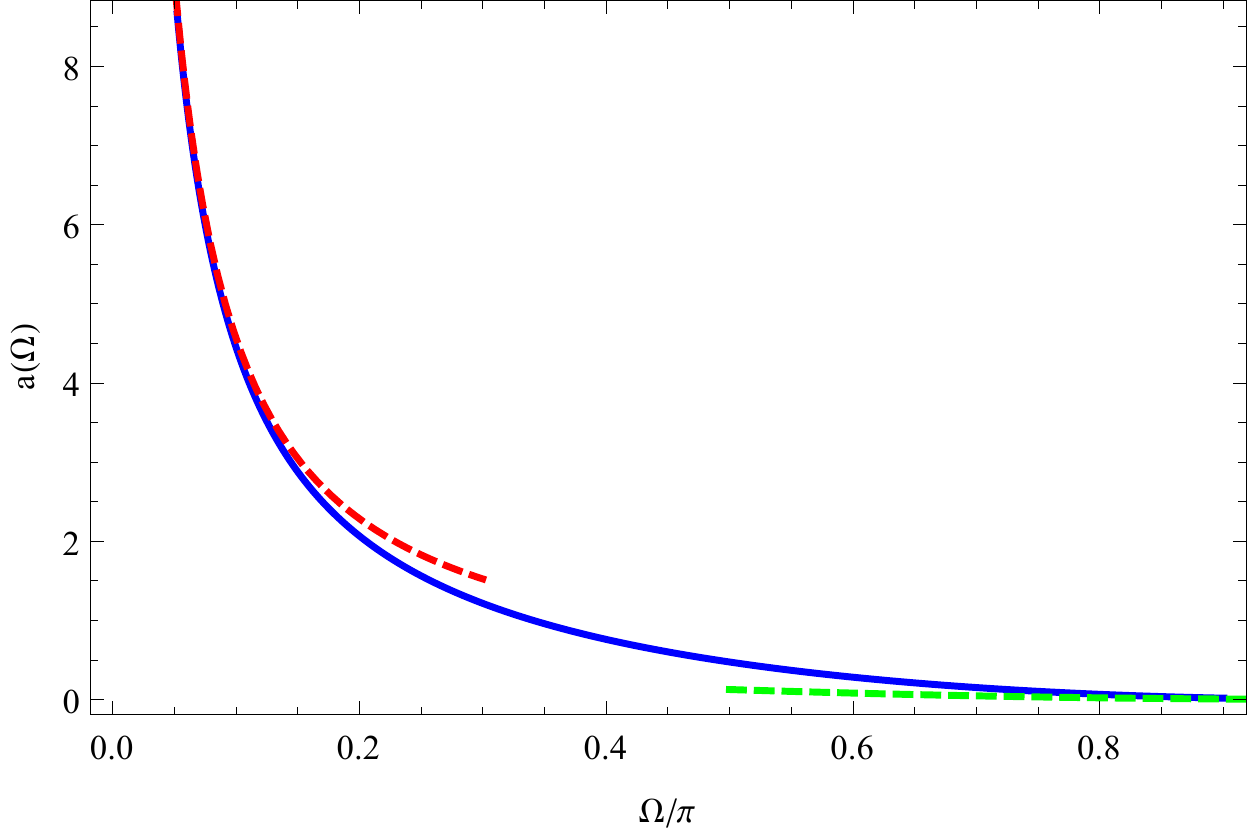}
  \includegraphics[width=210pt]{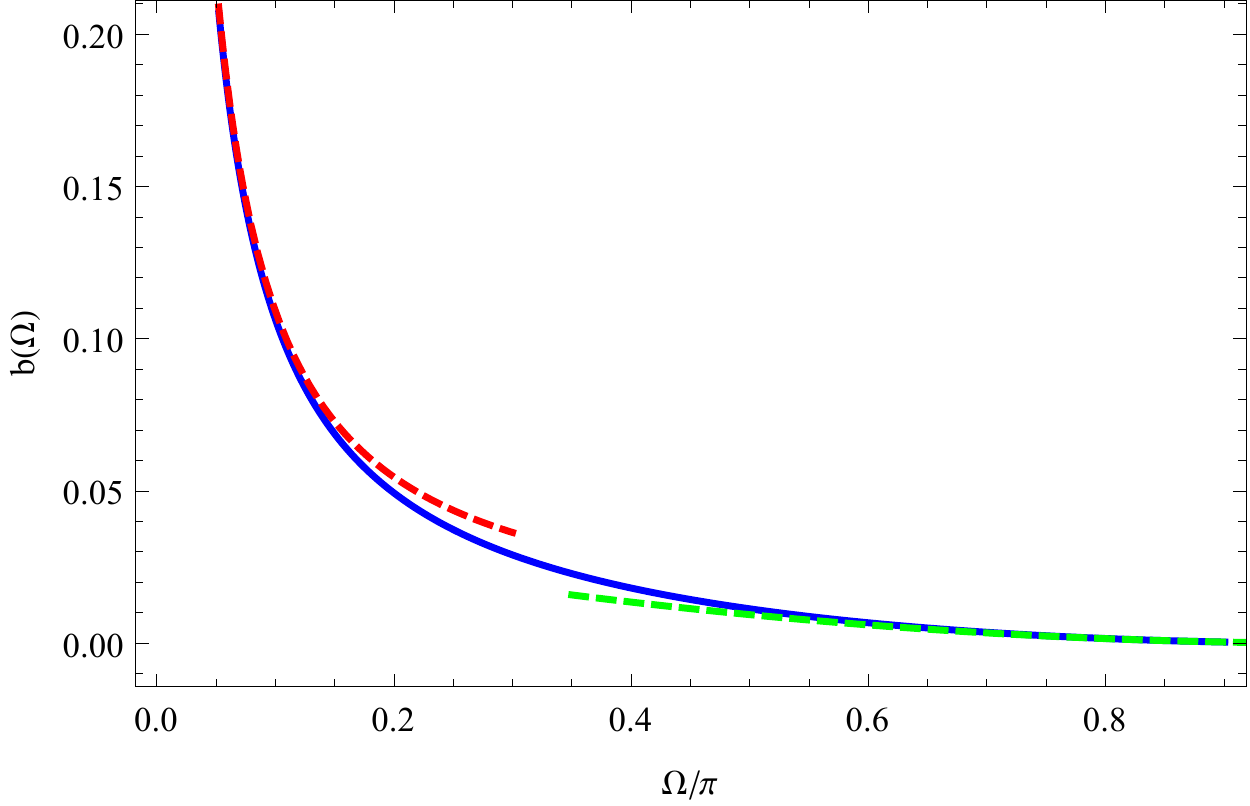}
  \caption{The plots for the function $a(\Omega)$ (left panel) and $b(\Omega)$ (right panel). For small opening angle $\Omega\rightarrow 0$, one has to leading order $a(\Omega)=2\Gamma{(\fft{3}{4})}^4/(\pi\Omega)$ and $b(\Omega)=3\Gamma{(\fft{3}{4})}^4/\big(20\pi^2\Omega\big)$ (shown in dashed red) whilst in the smooth limit
  $\Omega\rightarrow \pi$, $a(\Omega)=(\pi-\Omega)^2/2\pi$ and $b(\Omega)=3(\pi-\Omega)^2/80\pi^2$ (show in dashed green). In both limits, $b(\Omega)/a(\Omega)=3/40\pi$.}
  \label{cuspab}
\end{figure}

The induced metric on the surface is given by
\be ds^2_\Sigma=\fft{\ell^2_{AdS}}{r^2}\Big(1+\fft{1}{\varphi^2} \Big)dr^2+\fft{\ell^2_{AdS}}{\varphi^2}\big(1+\dot{\varphi}^2\big)d\theta^2+\fft{2\ell^2_{AdS}\dot{\varphi}}{r\varphi}drd\theta\,.\ee
Evaluation of the area functional yields
\be A_\Sigma=2\ell^2_{AdS}\int^R_{\delta/\varphi_0}\fft{dr}{r}\int_0^{\Omega/2-\varepsilon}d\theta\,\fft{\sqrt{1+\varphi^2+\dot{\varphi}^2}}{\varphi^2} \,, \ee
where $\varphi_0\equiv \varphi(0)$ stands for the maximum value of the function. Here the angular cutoff $\varepsilon$ is defined such that $\xi=\delta\,,\varphi(\Omega/2-\varepsilon)=\delta/r$. Variation of the area functional determines the minimal surface as \cite{Hirata:2006jx}
\be\label{cuspeq} \fft{1+\varphi^2}{\varphi^2\sqrt{1+\varphi^2+\dot{\varphi}^2}}=\fft{\sqrt{1+\varphi_0^2}}{\varphi_0^2} \,.\ee
To compute the area of the minimal surface, it is more convenient to introduce a variable $\mu=\sqrt{1/\varphi^2-1/\varphi_0^2}$. One has \cite{Bueno:2015rda,Bueno:2015xda}
\bea\label{cuspRT}
\mc{S}&=&2\int^R_{\delta/\varphi_0}\fft{dr}{r}\int_0^{\sqrt{r^2/\delta^2-1/\varphi_0^2}}d\mu\,
\sqrt{\fft{1+\varphi_0^2(1+\mu^2)}{2+\varphi_0^2(1+\mu^2)}}\nn\\
&=& \fft{2R}{\delta}-a(\Omega)\log{\Big( \fft{R}{\delta}\Big)}+\cdots \,,
\eea
where dots stands for constant terms, which are now regulator dependent. Instead, physical information about the boundary theories is encoded in the function $a(\Omega)$, given by \cite{Bueno:2015rda,Bueno:2015xda}
\be\label{functiona} a(\Omega)=2\int_0^\infty d\mu\,\Big[1- \sqrt{\ft{1+\varphi_0^2(1+\mu^2)}{2+\varphi_0^2(1+\mu^2)}}\,\Big] \,.\ee
This is an implicit function of the opening angle $\Omega$ through the dependence of $\varphi_0$ on $\Omega$. One has
\be \Omega=\int_{-\Omega/2}^{+\Omega/2}d\theta=\int_0^{\varphi_0}d\varphi\,
\fft{2\varphi^2\sqrt{1+\varphi_0^2}}{\sqrt{1+\varphi^2}\sqrt{(\varphi_0^2-\varphi^2)\big(\varphi_0^2+(1+\varphi_0^2)\varphi^2\big)}}\,. \ee
Full numerical solution of the function $a(\Omega)$ was presented in \cite{Bueno:2015rda,Bueno:2015xda} and we reproduce it in the left panel of Fig.\ref{cuspab}. Moreover, careful examination of the function for both the small opening angle $\Omega\rightarrow 0$ and the smooth limit $\Omega\rightarrow \pi$, one finds to leading order \cite{Bueno:2015rda,Bueno:2015xda}
\bea
&&\Omega\rightarrow 0\,,\quad a(\Omega)=2\Gamma^4{\big(\ft{3}{4}\big)}/(\pi\Omega)+\cdots \,,\nn\\
&&\Omega\rightarrow \pi\,,\quad  a(\Omega)=(\pi-\Omega)^2/2\pi+\cdots \,.
\eea
It was established in \cite{Bueno:2015rda,Bueno:2015xda} that some universal information about the boundary theories can be read off from the above asymptotic expansions.

Return to the bulk entanglement entropy. We deduce the scalar curvature of the RT surface $\Sigma$ as
\bea
R_\Sigma &=&-\fft{2}{\ell_{AdS}^2}\,\fft{\dot{\varphi}^4+(1+2\varphi^2)\dot{\varphi}^2-\varphi(1+\varphi^2)\ddot{\varphi}}{(1+\varphi^2+\dot{\varphi}^2)^2} \nn\\
&=&-\fft{2}{\ell_{AdS}^2}\,\fft{\varphi_0^4\,(1+2\varphi^2)+(1+\varphi_0^2+\varphi_0^4)\varphi^4}{\varphi_0^4\,(1+\varphi^2)^2}\,,
\eea
where in the second line we have adopted the relation (\ref{cuspeq}). Using these results and evaluation of the one-loop bulk entanglement entropy yields
\be\label{Sqcusp1} s_q=\fft{1}{180\pi}\int^R_{\delta/\varphi_0}\fft{dr}{r}\int_0^{\sqrt{r^2/\delta^2-1/\varphi_0^2}}d\mu\,Y_\Omega(\mu)\,,\ee
 where
\be Y_\Omega(\mu)=\fft{2(1+\varphi_0^2)+29\big(1+\varphi_0^2(1+\mu^2)\big)^2}{\sqrt{2+\varphi_0^2(1+\mu^2)}\big(1+\varphi_0^2(1+\mu^2)\big)^{3/2}}\,.  \ee
The subscript of the function $Y_\Omega(\mu) $ reminds us that it depends on the opening angle $\Omega$ implicitly. Note that the right hand side of the equation (\ref{Sqcusp1}) diverges in the short distance limit $\delta\rightarrow 0$ because of $Y_\Omega(\mu)\rightarrow 29$ as $\mu\rightarrow \infty$. To isolate this divergence, we treat (\ref{Sqcusp1}) as
\bea\label{Sqcusp2}
s_q&=&\fft{1}{180\pi}\int^R_{\delta/\varphi_0}\fft{dr}{r}\int_0^{\infty}d\mu\,\Big[Y_\Omega(\mu)-29\Big]\nn\\
&&\qquad\quad+\fft{29}{180\pi}\int^R_{\delta/\varphi_0}\fft{dr}{r}\,\sqrt{\fft{r^2}{\delta^2}-\fft{1}{\varphi_0^2}}  \nn\\
   &=&\fft{29R}{180\pi \delta}-b(\Omega)\log{\Big(\fft{R}{\delta} \Big)}+\cdots\,,
\eea
where again the dots stands for constant terms which are regulator dependent. The function $b(\Omega)$ is defined as
\be\label{functionb} b(\Omega)=\fft{1}{180\pi}\int_0^\infty d\mu\,\Big[29-Y_\Omega(\mu)\Big] \,.\ee
It is interesting to notice that the result (\ref{Sqcusp2}) takes a similar form to (\ref{cuspRT}) for the leading order entanglement entropy. Full numerical solution of the function $b(\Omega)$ is shown in the right panel of Fig.\ref{cuspab}. At first sight, we find that it behaves very similar to the function $a(\Omega)$. In addition, examination of the function for a small opening angle as well as the smooth limit tells us that in both cases, their ratio is a constant
\be\label{ratio} \fft{b(\Omega)}{a(\Omega)}=\fft{3}{40\pi} \,.\ee
This motives us to study the function $b(\Omega)$ more carefully. As a matter of fact, we can analytically show that
the ratio is valid for an arbitrary opening angle $\Omega$, despite that the integrated function in (\ref{functionb}) looks quite different from that in (\ref{functiona}). The proof is straightforward by making use of a mathematical identity
\be
\int_0^\infty d\mu \,f(\mu)=\mu \Big( 1-\sqrt{\ft{2+\varphi_0^2(1+\mu^2)}{1+\varphi_0^2(1+\mu^2)}}\, \Big)\Big|_0^\infty=0\,,
\ee
where
\be f(\mu)=1-\sqrt{\ft{1+\varphi_0^2(1+\mu^2)}{2+\varphi_0^2(1+\mu^2)}}-\ft{1+\varphi_0^2}{\sqrt{2+\varphi_0^2(1+\mu^2)}\big(1+\varphi_0^2(1+\mu^2)\big)^{3/2}}\,. \ee
This strongly implies that the same information about holographic $\mm{CFT}_3$ can be read off from both the bulk entanglement entropy and the leading entanglement entropy of singular shapes, except for a constant ratio. It is remarkable that the result does not depend on the opening angle of the cone in the boundary.

\section{Shape dependence of bulk entanglement}\label{shape}
Having studied the bulk entanglement entropy for various shapes of boundary subregions, we would like to further investigate its shape dependence to see whether the same physical information can be extracted in this case as that in the leading entanglement entropy. We first consider the $D=4$ dimension and then extend the discussions to the $D=6$ dimension.

\subsection{Deformed circle in the $D=4$ dimension}

 Consider a generally smooth entangling surface in the boundary, obtained by slightly deforming a circle
\be r(\phi)/R=1+\hat\epsilon\sum_{\ell=1}^\infty \Big(a_\ell \fft{\cos{(\ell\phi)}}{\sqrt{\pi}}+b_\ell \fft{\sin{(\ell\phi)}}{\sqrt{\pi}} \Big) \,,\ee
where $\hat\epsilon$ is a small parameter. We have properly chosen the expansion coefficients $a_\ell\,,b_\ell$ so that the Fourier modes are normalized to unity.

 As shown in \cite{Allais:2014ata}, the bulk minimal surface in this case becomes a deformed hemisphere and the lowest order correction to the leading entanglement entropy appears at $\hat{\epsilon}^2$. We shall briefly review these results in the following and then calculate the one-loop bulk entanglement entropy. We will show that the lowest order correction to the bulk entanglement appears at $\hat{\epsilon}^2$ as well.

 For later convenience, we introduce new coordinates $(\rho\,,\theta)$, under which the bulk metric reads
 \be ds^2=\fft{\ell^2_{AdS}}{\rho^2\cos^2\theta}\Big(dt_E^2+d\rho^2+\rho^2\big(d\theta^2+\sin\theta^2 d\phi^2\big) \Big) \,,\ee
 where
\be \rho=\sqrt{\xi^2+r^2}\,,\quad \theta=\mm{arctan}\big(\fft{r}{\xi}\big) \,,\ee
where $\theta\in [0\,,\pi/2]$ and $\theta=\pi/2$ corresponds to the asymptotic AdS boundary. Notice that the constant $\rho$ slices describe bulk hemispheres. Hence, a deformed hemisphere can be parameterized as
\be \rho/R=1+\hat\epsilon\,\rho_1(\theta\,,\phi)+O(\hat{\epsilon}^2)  \,,\ee
where the higher order terms $O(\hat{\epsilon}^2)$ do not contribute to the leading entanglement entropy \cite{Allais:2014ata} as well as the bulk entanglement at $\hat{\epsilon}^2$ order, as will be shown later.

The induced metric of the bulk surface is given by
\be ds^2_\Sigma=\fft{\ell^2_{AdS}}{\rho^2\cos^2\theta}\Big[\big(\rho_\theta^2+\rho^2 \big)d\theta^2+2\rho_\theta \rho_\phi d\theta d\phi+\big(\rho_\phi^2+\rho^2\sin^2\theta \big)d\phi^2 \Big] \,,\ee
where $\rho_\theta\equiv \partial_\theta\rho\,,\rho_\phi=\partial_\phi \rho$. The deformations $\rho_1(\theta\,,\phi)$ can be solved analytically by minimizing the area functional
\be A_\Sigma=\ell^2_{AdS}\int_0^{\pi/2-\varepsilon}d\theta \int_0^{2\pi}d\phi\,\fft{1}{\rho\cos^2\theta}\sqrt{\rho_\phi^2+\big(\rho^2_\theta+\rho^2 \big)\sin^2\theta} \,,\ee
where the angular cutoff $\varepsilon$ is related to the short distance cutoff $\delta$ at the asymptotic boundary as $\delta=R \cos(\pi/2-\varepsilon)$.
Then straightforward calculations lead to \cite{Allais:2014ata,Mezei:2014zla}
\be \rho_1(\theta\,,\phi)= \sum_{\ell=1}^\infty  \Big(a_\ell \fft{\cos{(\ell\phi)}}{\sqrt{\pi}}+b_\ell \fft{\sin{(\ell\phi)}}{\sqrt{\pi}} \Big) \tan^\ell{\big(\fft{\theta}{2}\big)}(1+\ell\cos\theta)\,.\ee
Evaluation of the area functional gives
\be\label{smoothRT} \mc{S}=\fft{2\pi R}{\delta}\Big(1+\hat{\epsilon}^2\,\sum_\ell\ft{\ell^2(a_\ell^2+b_\ell^2)}{4} \Big)-\Big(2\pi+\hat{\epsilon}^2\,\sum_\ell\fft{1}{2}\ell(\ell^2-1)(a_\ell^2+b_\ell^2) \Big) \,.\ee

To compute the one-loop bulk entanglement, we deduce the scalar curvature for the deformed hemisphere
\be R_\Sigma=-2\ell^{-2}_{AdS}\Big( 1+\hat{\epsilon}^2\ell^2(\ell^2-1)^2(a_\ell^2+b_\ell^2)\cot^4\theta \,\tan^{2\ell}{\big(\fft{\theta}{2}\big)} \Big) \,. \ee
It is clear that the lowest order corrections appears at $\epsilon^2$. We obtain
\be\label{smoothsq} s_q=-\fft{29R}{180\delta}\Big(1+\hat{\epsilon}^2 \sum_\ell\ft{\ell^2(a_\ell^2+b_\ell^2)}{4}\Big)+\Big(\fft{29}{180}+\hat{\epsilon}^2\,\sum_\ell\fft{3}{80\pi}\ell(\ell^2-1)(a_\ell^2+b_\ell^2) \Big)\,.\ee
It is interesting to observe that the bulk entanglement shares the same shape dependence with the leading entanglement entropy in the boundary, except for a constant factor $\fft{3}{40\pi}$.
This is highly non-trivial since in the boundary, the universal piece of the leading entanglement entropy is a constant, encoding nonlocal information about the underlying theories while the bulk entanglement just contains local information about $O(1)$ degrees of freedoms in the boundary. This local/nonlocal duality implies that for holographic $\mm{CFT}_3$, the same physical information can be extracted from the entanglement entropy at either the leading order $O(N^2)$ or the subleading order $O(1)$ (this is said in the sense that interactions between the $O(1)$ and $O(N^2)$ degrees of freedoms is ignored since we have not included backreaction effects of bulk fluctuations ).

Last but not least, we realize that the particular value of the ratio $\fft{3}{40\pi}$ is the same as (\ref{ratio}) for the entanglement entropy of singular shapes. This is easily explained for the smooth limit $\Omega\rightarrow \pi$. It was shown in \cite{Bueno:2015lza} that the universal term of the leading entanglement entropy (\ref{cuspRT})
for singular shapes can be derived from that for the smoothly deformed spheres (\ref{smoothRT}). It is clear that this will also be the case for the bulk entanglement due to (\ref{smoothsq}) and hence the coincidence of the ratio $\fft{3}{40\pi}$ is not a surprise.

\subsection{ Deformed spheres in the $D=6$ dimension}\label{sec5}
Inspired by the interesting results in four dimension, we would like to investigate the shape dependence of bulk entanglement in higher dimensions to see whether the local/nonlocal duality is still valid. We focus on the $D=6$ dimension in this subsection. The relevant heat kernel coefficient is $a_3$, given by \cite{Vassilevich:2003xt}
\bea
a_3&=&\fft{1}{7!}\int d^Dx\,\sqrt{g}\,\,\Big[\fft{35}{9}R^3-\fft{14}{3}R R_{\mu\nu}R^{\mu\nu}+\fft{14}{3}R R_{\mu\nu\rho\sigma}R^{\mu\nu\rho\sigma}\nn\\
&&\qquad\qquad\qquad\quad -\fft{208}{9}R_{\mu\lambda}R^\mu_{\,\,\,\, \rho}R^{\lambda\rho}-\fft{64}{3}R_{\mu\nu}R_{\lambda\rho}R^{\mu\lambda\nu\rho}-\fft{16}{3}R_{\mu\lambda}R^\mu_{\,\,\,\, \alpha\beta\rho}R^{\lambda\alpha\beta\rho}\nn\\
&&\qquad\qquad\qquad\quad-\fft{44}{9}R_{\mu\nu}^{\quad\lambda\rho}R_{\lambda\rho}^{\quad\alpha\beta}
R_{\alpha\beta}^{\quad\mu\nu}-\fft{80}{9}R_{\mu\,\,\,\,\nu}^{\,\,\,\,\lambda\,\,\,\,\rho}
R_{\lambda\,\,\,\,\rho}^{\,\,\,\,\alpha\,\,\,\,\beta}R^{\,\,\,\,\mu\,\,\,\,\nu}_{\alpha\,\,\,\,\beta\,\,\,\,}+der
\Big]\,,
\eea
where $der$ stands for derivative terms of curvatures, given by
\bea
der&=&18\Box\Box R+17\big(\nabla R \big)^2-2\nabla_\sigma R_{\mu\nu}\nabla^\sigma R^{\mu\nu}-4\nabla_\sigma R_{\mu\nu}\nabla^\mu R^{\nu\sigma} \nn\\
&&+9\nabla_\sigma R_{\mu\nu\lambda\rho}\nabla^\sigma R^{\mu\nu\lambda\rho}+28R\Box R-8R_{\mu\nu}\Box R^{\mu\nu} \\
&&+24 R^\mu_{\,\,\,\,\lambda}\nabla_\mu\nabla_\rho R^{\lambda\rho}+12 R_{\mu\nu\rho\sigma}\Box R^{\mu\nu\rho\sigma} \,.\nn
\eea
However, since total derivative terms do not contribute to $(\partial_n a_{3\,,n})_{n=1}$ \cite{Dong:2015zba}, we can do an integration by parts and drop all these terms. This simplifies $der$ as
\bea
der&=&-18\big(\nabla R \big)^2+6\nabla_\sigma R_{\mu\nu}\nabla^\sigma R^{\mu\nu}-3 \nabla_\sigma R_{\mu\nu\lambda\rho}\nabla^\sigma R^{\mu\nu\lambda\rho} \nn\\
&&+4\big(R_{\mu\lambda}R^\mu_{\,\,\,\,\rho}R^{\lambda\rho}-R_{\mu\nu}R_{\lambda\rho}R^{\mu\lambda\nu\rho} \big) \,.
\eea
The coefficient $a_3$ can be reorganized as
\be  a_3=a_3^{(1)}+a_3^{(2)}\,,\ee
where
\bea\label{a3final}
&&a_3^{(1)}=\fft{1}{7!}\int d^Dx\,\sqrt{g}\,\,\Big[\,\fft{35}{9}R^3-\fft{14}{3}R R_{\mu\nu}R^{\mu\nu}+\fft{14}{3}R R_{\mu\nu\rho\sigma}R^{\mu\nu\rho\sigma} -\fft{172}{9}R_{\mu\lambda}R^\mu_{\,\,\,\, \rho}R^{\lambda\rho}\nn\\
&&\qquad\qquad\qquad\qquad\qquad-\fft{76}{3}R_{\mu\nu}R_{\lambda\rho}R^{\mu\lambda\nu\rho}-\fft{16}{3}R_{\mu\lambda}R^\mu_{\,\,\,\, \alpha\beta\rho}R^{\lambda\alpha\beta\rho}\nn\\
&&\qquad\qquad\qquad\qquad\qquad-\fft{44}{9}R_{\mu\nu}^{\quad\lambda\rho}R_{\lambda\rho}^{\quad\alpha\beta}
R_{\alpha\beta}^{\quad\mu\nu}-\fft{80}{9}R_{\mu\,\,\,\,\nu}^{\,\,\,\,\lambda\,\,\,\,\rho}
R_{\lambda\,\,\,\,\rho}^{\,\,\,\,\alpha\,\,\,\,\beta}R^{\,\,\,\,\mu\,\,\,\,\nu}_{\alpha\,\,\,\,\beta\,\,\,\,}\Big]\,,\\
&&a_3^{(2)}=\int d^{D}x\,\sqrt{g}\,\Big( -\fft{1}{280} \big(\nabla R\big)^2  +\fft{1}{840} \nabla_\sigma R_{\mu\nu}\nabla^\sigma R^{\mu\nu}-\fft{1}{1680} \nabla_\sigma R_{\mu\nu\lambda\rho}\nabla^\sigma R^{\mu\nu\lambda\rho}\Big)\,,\nn
\eea
where the first term involves third order curvature polynomials whilst the second term contains first order derivatives of curvatures. Since the contributions of the above two terms are highly different, we shall deal with them separately.

According to (\ref{dong}), the cubic Riemannian term $a_3^{(1)}$ can be evaluated as
\be \partial_n a_{3\,,n}^{(1)}\Big|_{n=1}=\mm{Wald}+\mm{Anomaly} \,, \ee
where the shorthand notations on the r.h.s are specified in (\ref{wald}) and (\ref{anomaly}). We quote them as follows
\bea\label{a3eva}
&&\mm{Wald}=-2\pi\int_\Sigma E^{\mu\nu\rho\sigma}\varepsilon_{\mu\nu}\varepsilon_{\rho\sigma}\,,\nn\\
&& \mm{Anomaly}=2\pi \int_\Sigma\,\sum_\alpha \Big(\fft{\partial^2 L}{\partial R_{izjz} \partial R_{k\bar{z}l \bar{z}}}\Big)_\alpha\fft{8K_{zij} K_{\bar{z}kl} }{q_\alpha+1} \,,
\eea
where $E^{\mu\nu\rho\sigma}=\partial L/\partial R_{\mu\nu\rho\sigma}$. To proceed, we need derive the tensor $E^{\mu\nu\rho\sigma}=\partial L/\partial R_{\mu\nu\rho\sigma}$ for each of the cubic curvature polynomials as well as the second order derives $\fft{\partial^2 L}{\partial R_{izjz}\partial R_{k\bar{z}\ell \bar{z}}}$. The calculations are straightforward but a bit lengthy. We refer the readers to Appendix B for details. The results for general cases are quite involved but will be greatly simplified for minimal area surfaces in AdS vacuum. Here we just present the final result
\be\label{a31} \partial_n a_{3\,,n}^{(1)}\Big|_{n=1}=-\fft{46\pi}{2835\ell^2_{AdS}}\int_\Sigma \big(273\ell^{-2}_{AdS}+K_{aij}K^{aij} \big) \,.\ee

The second term $a_3^{(2)}$ in (\ref{a3final}), as a gravitational action, its contribution to holographic entanglement entropy was studied carefully in \cite{Miao:2014nxa} for general coupling constants
\be\label{derivativeaction} I=\int d^{D}x\,\sqrt{g}\,\Big( \lambda_1 \nabla_\mu R \nabla^\mu R  +\lambda_2 \nabla_\sigma R_{\mu\nu}\nabla^\sigma R^{\mu\nu}+\lambda_3 \nabla_\sigma R_{\mu\nu\lambda\rho}\nabla^\sigma R^{\mu\nu\lambda\rho}\Big)\,.\ee
In our case (\ref{a3final}), $\lambda_1=-1/280\,,\lambda_2=1/840\,,\lambda_3=-1/1680$. One has
\be \partial_n I[\hat{\mc{M}}_n]\Big|_{n=1}=\mm{\underline{Wald}}+\mm{\underline{Anomaly}}\,,\ee
where the generalised Wald entropy term is given by (\ref{waldbar}). For this particular action
(\ref{derivativeaction}), one has \cite{Miao:2014nxa}
\bea
\underline{\mm{Wald}}&=&4\pi\int_\Sigma 2\lambda_1\, \Box R+\lambda_2\, n_{\mu\nu}\Box R^{\mu\nu}+\lambda_3\, \varepsilon_{\mu\nu}\varepsilon_{\rho\sigma}\Box R^{\mu\nu\rho\sigma}  \nn\\
&&\qquad\quad+\big(\, \fft12 \lambda_2\, \nabla^\alpha R^{\mu\nu}K_\beta+2\lambda_3\, \nabla_\alpha R^{\mu\rho\nu\sigma} K_{\beta\rho\sigma} \big) \big(n^\beta_{\,\,\,\,\mu}n_{\alpha\nu}-\varepsilon^\beta_{\,\,\,\,\mu}\varepsilon_{\alpha\nu} \big)\,.
\eea
However, this term will vanish for RT surfaces in AdS vacuum, where the Riemann curvature takes a particularly simple form
\be R_{\mu\lambda\nu\rho}=-\ell^{-2}_{AdS}\big(g_{\mu\nu}g_{\lambda\rho}-g_{\mu\rho}g_{\nu\lambda}\big) \,.\ee
Derivation of the anomaly terms is much more involved  and the results have lengthy expressions as well, see the Appendix A of \cite{Miao:2014nxa}. Fortunately, in this subsection, we focus on deformed hemispheres, which just has small extrinsic curvatures $K\sim \hat\epsilon$. We have known that in this case, the lowest order correction of the smooth deformations to the universal terms of entanglement entropy appears at the quadratic order $O(K^2)$. Thus, we can drop all the higher order terms presented in \cite{Miao:2014nxa}. We find
\be
\nabla_\mu R \nabla^\mu R:\qquad \underline{\mathrm Anomaly}=O(K^4)\,.
\ee
For $\nabla_\sigma R_{\mu\nu}\nabla^\sigma R^{\mu\nu}$,
\bea\label{anomaly2}
\underline{\mathrm Anomaly}&=&-4\pi\int_\Sigma  \Big (K_{aij}K_b^{\,\,\,\,ij}Q_{cd}-\fft12 Q_{ab}Q_{cd}\nn\\
  &&\qquad\qquad\qquad-\fft18\big(R_{ai}R_{b}^{\,\,\,\,i}+R_{ci}R_d^{\,\,\,\,i} \big)\Big)
 \big(n^{ac}n^{bd}-\varepsilon^{ac}\varepsilon^{bd} \big)+O(K^4)\,,
\eea
and for $\nabla_\sigma R_{\mu\nu\alpha\beta}\nabla^\sigma R^{\mu\nu\alpha\beta}$,
\bea
 \underline{\mathrm Anomaly}&=&4\pi\int_\Sigma\Big[\,2 \big(Q_{abij}Q_{cd}^{\quad ij}-3P_{abcij}K_d^{\,\,\,\,ij}\big)\big(n^{ac}n^{bd}-\varepsilon^{ac}\varepsilon^{bd} \big)\nn\\
 &&\qquad\qquad+4 K_{a\ell}^{\quad i}K^{a\ell j}Q^b_{\,\,\,\,bij} +\big(Q^b_{\,\,\,\,b}-80T_1 \big) K_{aij}K^{aij}\nn\\
 &&\qquad\qquad\qquad\qquad\qquad+2\mc{D}_{\ell}K_{aij}\mc{D}^\ell K^{aij}+2R_{aijk}R^{aijk}\Big]+O(K^4)\,.
\eea
However, here we have not considered the relation between $Q_{abij}\,,P_{abcij}$ and the extrinsic curvatures. By comparing the Riemann tensor in the metric expansion (\ref{metricexpansion}) with that for pure AdS, one finds \cite{Miao:2014nxa}
\bea
&&T_1=-\ft{1}{12\ell^2_{AdS}}\,,\quad V_i=0\,,\quad Q_{zzij}=K_{zi\ell}K_{zj}^{\quad\ell}\,,\nn\\
&&Q_{\bar{z}\bar{z}ij}=K_{\bar{z}i\ell}K_{\bar{z}j}^{\quad\ell}\,,\quad Q_{z\bar{z}ij}=\ft{1}{2\ell^2_{AdS}}g_{ij}+K_{zi\ell}K_{\bar{z}j}^{\quad\ell}\,,\nn\\
&& P_{zz\bar{z}ij}=\ft{4}{9\ell^2_{AdS}}K_{zij}+O(K^3)\,,\quad P_{\bar{z}\bar{z}zij}=\ft{4}{9\ell^2_{AdS}}K_{\bar{z}ij}+O(K^3)\,.
\eea
Using these relations, we find that remarkably all the terms in (\ref{anomaly2}) are of quartic order $O(K^4)$ and hence are irrelevant for our discussions. Finally, the anomaly terms for the derivative action (\ref{derivativeaction}) greatly simplify to
\be
 \underline{\mathrm Anomaly}=8\pi\lambda_3\int_\Sigma \big(D \ell^{-2}_{AdS}\,K_{aij}K^{aij}+\mc{D}_{\ell}K_{aij}\mc{D}^\ell K^{aij} \big)\,.
\ee
Combing all the results together, we deduce
\be\label{a3der} \partial_n a_{3\,,n}\Big|_{n=1}=-\ft{598\pi}{135}\,\mc{S}-\int_\Sigma \Big(\ft{127\pi}{2835\ell^2_{AdS}}\,K_{aij}K^{aij}+\ft{\pi}{210}\, \mc{D}_{\ell}K_{aij}\mc{D}^\ell K^{aij}\Big)\,.\ee

To proceed, we need construct the adapted coordinates around the deformed hemispheres $\rho_\Sigma=1+\epsilon \rho_1(\Omega_d)$ and read off the extrinsic curvatures. This is achieved by constructing geodesics emanating from the surface \cite{Mezei:2014zla}. We denote the affine parameter of the geodesics by $s$ and $s=0$ at the surface. We set $\partial_s\cdot \partial_t=\sin\tau$ and the starting point on the surface is $\omega_d=(\theta\,,\omega_{d-1})$. The geodesics can be constructed by solving the geodesic equation in a power series of $s$ and hence covering a small neighborhood of the surface using the new coordinates $(\tau\,,s\,,\omega_d)$. We present the linear in $s$ peace as
\bea
&&t=s \sin\tau \cos\theta \,\rho_\Sigma(\omega_d)+\cdots\,,\nn\\
&&\rho=\rho_\Sigma(\omega_d)+s\cos\tau\cos\theta\, \rho_\Sigma(\omega_d)+\cdots\,,\nn\\
&&\Omega_d=\omega_d-s\cos\tau\cos\theta\,\partial_{\omega_d}\rho_\Sigma(\omega_d)+\cdots\,,
\eea
where $\partial_{\omega_d}=\tilde{g}^{ij}\partial_j$, where $\tilde{g}_{ij}$ is the metric of unit $\mathbf{S}_d$. Finally, we change the variables to $z=s e^{i\tau}\,,\bar{z}=s e^{-i\tau}$. The extrinsic curvatures can be read off straightforwardly from the construction for each dimension. We find that the results, valid to general dimensions, can be expressed compactly as
\be\label{kzij} K_{zij}=K_{\bar{z}ij}=\fft12 \Big(\mc{D}_i m_j+\gamma^{(0)}_{ij}\,\fft{\epsilon\rho_1}{\cos\theta}\,\Big)+O(\epsilon^2) \,,\ee
where
\be m_i=-\ell_{AdS}^2\,\partial_i\Big(  \fft{\epsilon\rho_1}{\cos\theta}\Big)\,,\quad \gamma^{(0)}_{ij}=\fft{\ell^2_{AdS}}{\cos\theta^2}\,\tilde{g}_{ij} \,.\ee

\subsubsection{Explicit calculations}
After preparing so much, we are ready to perform explicit calculations for the one-loop bulk entanglement entropy for deformed hemispheres. In the boundary, the entangling surface is described by \cite{Mezei:2014zla}
\be r(\Omega_{d-1})/R=1+\hat\epsilon\sum_{\ell\,,m_1\,\cdots\,,m_{d-2}} a_{\ell\,,m_1\,\cdots\,,m_{d-2}}Y_{\ell\,,m_1\,\cdots\,,m_{d-2}}(\Omega_{d-1}) \,,\ee
where $\Omega_{d-1}$ are the angular coordinates on $\mathbf{S}_{d-1}$ and $Y_{\ell\,,m_1\,\cdots\,,m_{d-2}}(\Omega_{d-1})$ are hyperspherical harmonics\footnote{They are eigenfunctions of the Laplacian on $\mathbf{S}_{d-1}$:
\be \Delta Y_{\ell\,,m_1\,\cdots\,,m_{d-2}}(\Omega_{d-1})=-\ell(\ell+d-2)Y_{\ell\,,m_1\,\cdots\,,m_{d-2}}(\Omega_{d-1})\,.\ee}.
We follow \cite{Mezei:2014zla} and normalize the hyperspherical harmonics as
\be \int d\Omega_{d-1}Y_{\ell\,,m_1\,\cdots\,,m_{d-2}}Y_{\ell'\,,m'_1\,\cdots\,,m'_{d-2}}=\delta_{\ell\ell'}\delta_{m_1m_1'}\cdots \delta_{m_{d-2}m'_{d-2}}\,. \ee
As in the four dimensional case, the deformed hemisphere can be most easily solved under the $(\rho\,,\theta)$ coordinates, where the bulk metric reads
\be ds^2=\fft{\ell^2_{AdS}}{\rho^2\cos^2\theta}\Big(dt_E^2+d\rho^2+\rho^2\big(d\theta^2+\sin\theta^2 d\Omega_{d-1}^2\big) \Big) \,.\ee
The deformed hemispheres can be parameterized as
\be \rho/R=1+\hat\epsilon\,\rho_1(\theta\,,\Omega_{d-1})+O(\hat{\epsilon}^2)  \,,\ee
where again the higher order terms $O(\hat{\epsilon}^2)$ do not contribute to entanglement entropy at $\hat{\epsilon}^2$ order. It follows that the deformation $\rho_1(\theta\,,\Omega_{d-1})$ can be solved analytically by minimizing the area functional
\be A_\Sigma=\ell^d_{AdS}\int_0^{\pi/2-\varepsilon}d\theta \int d\Omega_{d-1}\,\fft{(\sin\theta)^{d-2}}{\rho(\cos\theta)^d}\sqrt{\rho_\Omega^2+\big(\rho^2_\theta+\rho^2 \big)\sin^2\theta} \,,\ee
where $\rho_\theta\equiv \partial_\theta\rho\,,\rho_\Omega=\sqrt{\tilde{g}^{ij}}\partial_j\rho$, where $\tilde{g}_{ij}$ is the metric of unit $\mathbf{S}_{d-1}$. Again the angular cutoff $\varepsilon$ is related to the short distance cutoff $\delta$ at the asymptotic boundary as $\delta=R \cos(\pi/2-\varepsilon)$.

In the $D=6$ dimension, one finds \cite{Mezei:2014zla}
\be \rho_1(\theta\,,\Omega_3)= \sum_{\ell\,,m_1\,,m_2}a_{\ell\,,m_1\,,m_2} Y_{\ell\,,m_1\,,m_2}(\Omega_3)f(\theta)\,,\ee
where
\be f(\theta)=\tan^\ell{\big(\fft{\theta}{2}\big)}\fft{1+(\ell+1)\cos\theta+\ft{\ell(\ell+2)}{3}\cos^2\theta}{1+\cos\theta} \,.\ee
Evaluation of the area functional yields
\be \mc{S}=\fft{R^3}{\delta^3}-\fft{R}{\delta}+\fft{4\pi^2}{3}+\fft{\hat{\epsilon}^2}{18}\sum_{\ell\,,m_1\,,m_2} a^2_{\ell\,,m_1\,,m_2}\ell(\ell^2-1)(\ell+2)(\ell+3) \,.\ee

To proceed, we present explicit formulas for the three dimensional hyperspherical harmonics \cite{campos2020}
\be Y_{\ell\,,m_1\,,m_2}(\Omega_3)=P_{\mu\,,1/2}^\nu(\cos\phi_1)Y_{m_1\,,m_2}(\phi_2\,,\phi_3) \,,\ee
where $Y_{m_1\,,m_2}(\phi_2\,,\phi_3)$ is the usual spherical harmonics and $P_{\mu\,,1/2}^\nu(\cos\phi_1)$ is a hyper-Legendre function, given by
\be P_{\mu\,,1/2}^\nu(\cos\phi_1)=N_{\ell m_1}\big(\sin{\phi_1}\big)^{m_1}\,F\Big(\fft{m_1-\ell}{2}\,,\fft{m_1+\ell+2}{2}\,,\fft12\,,\big(\cos\phi_1\big)^2\Big) \,,\ee
where
\be \nu=\pm\sqrt{m_1(m_1+1)}\,,\quad \mu=\fft{-1\pm\sqrt{4\ell^2+8\ell+1}}{2} \,,\ee
and $N_{\ell m_1}$ is a normalization constant so that
\be \int_0^\pi d\phi_1\,\sin\phi_1^2\,P_{\mu\,,1/2}^\nu(\cos\phi_1)P_{\mu'\,,1/2}^{\nu}(\cos\phi_1)= \delta_{\ell \ell'} \,.\ee
According to (\ref{a3der}), the key elements to derive the one-loop bulk entanglement entropy are surface integrals of extrinsic curvatures and their spatial derivatives evaluated on the minimal surface. By straightforward calculations, we find
\bea
\int_\Sigma \ell^{-2}_{AdS}\,K_{aij}K^{aij}=-\fft{\hat{\epsilon}^2}{3}\sum_{\ell\,,m_1\,,m_2} a^2_{\ell\,,m_1\,,m_2}\ell(\ell^2-1)(\ell+2)(\ell+3) \,.
\eea
Again as in the four dimensional case, this term depends on the shape of entangling surface in the same manner as the leading entanglement entropy. However, for the derivative terms of extrinsic curvatures, the situation turns out to be much more complicated. To clarify this, we may set
\bea
\int_\Sigma \,\mc{D}_\ell K_{aij} \mc{D}^\ell K^{aij}\equiv \fft{\hat{\epsilon}^2}{18}\sum_{\ell\,,m_1\,,m_2}q_{\ell\,,m_1\,,m_2}\, a^2_{\ell\,,m_1\,,m_2}\ell(\ell^2-1)(\ell+2)(\ell+3)\,,
\eea
where the dependence on the shape of entangling surface is encoded in the functional relation $q_{\ell\,,m_1\,,m_2}$.
However, it is of great difficult to calculate this term for general eigenvalues analytically. Nevertheless, we can check some special cases to see whether it is a same constant for general eigenvalues. This is enough for our purpose. We present some low lying examples as follows
\bea
&&q_{\ell\,,m_1\,,0}=-24\,,\nn\\
&&q_{2\,,2\,,1}=\ft{-5193+2656\log{2}}{300}\simeq -11.1733\,,\nn\\
&& q_{2\,,2\,,2}=\ft{2(-481+1881\log{2})}{75}\simeq 21.9416\,,\nn\\
&&q_{3\,,1\,,1}=\ft{-7957+1772\log{2}}{375}\simeq -17.9433 \,,\nn\\
&&q_{4\,,2\,,1}=\ft{-155687+31488\log{2}}{6860}\simeq -19.5133\,.
\eea
It is clear that $q_{\ell\,,m_1\,,m_2}$ is no longer a same constant for general eigenvalues. Mathematically this is not hard to explain since the spatial derivatives of the hyperspherical harmonics generally mixes different Fourier modes of the entangling surface. It implies that in the $D=6$ dimension, the bulk entanglement entropy at $O(\hat{\epsilon}^2)$ order depends on the shape of entangling surface more strongly than the area of the RT surface. It may encode more universal information about the boundary theories at this order. We expect that in general this will also be the case for higher dimensions, since more higher order derivative terms of Riemann curvatures will appear in the heat kernel coefficient $a_{[D/2]}$ and hence more spatial derivative terms of extrinsic curvatures appear in the bulk entanglement entropy.


\section{Conclusions}

In this paper, we adopt the heat kernel method to study one-loop bulk entanglement entropy in diverse dimensions. A shortcoming of the method is it does not capture the cut-off independence piece of bulk entanglement entropy in odd dimensions. As a consequence, we focus on even dimensions in this paper.

We perform explicit calculations in the $D=4$ dimension for several different shapes of subregions in the boundary. In particular, for a cusp subregion, we find that the bulk entanglement entropy encodes the same universal information about the boundary theories as the leading entanglement entropy, up to a fixed proportional constant. Furthermore, we study the shape dependence of bulk entanglement by considering a smoothly deformed circle. We find that at leading order of the deformations, the bulk entanglement entropy shares the same shape dependence with the leading entanglement entropy in the boundary. This is interesting since the former just captures local information about $O(1)$ degrees of freedoms in the boundary whilst the latter encodes nonlocal information about $O(N^2)$ degrees of freedoms. The result establishes a local/nonlocal duality for shape dependence of entanglement entropy for holographic $\mm{CFT}_3$.

To see whether the same results hold for higher dimensions, we extend our investigations to the $D=6$ dimension. We find that the answer is no. The reason is in $D\geq 6$ dimensions, the heat kernel coefficient $a_{[D/2]}$, which captures the physical information about the underlying theories, contains covariant derivatives of Riemann curvatures. As a consequence, the one-loop bulk entanglement entropy will depend on spatial derivatives of extrinsic curvatures along the RT surfaces and hence depends on the shape of boundary subregions more strongly. This implies that in general the nice result in the $D=4$ dimension is not valid to general $D\geq 6$ even dimensions.

It is also interesting to extend our studies to odd spacetime dimensions. We leave this as a research direction in the near future.

\section*{Acknowledgments}
Z.Y. Fan was supported in part by the National Natural Science Foundations of China with Grant No. 11805041 and No. 11873025.

\appendix
\section{Curvatures for orbifold geometry}

Let us calculate curvatures for a product metric as
\be ds^2=G_{ab}dx^a dx^b+G_{ij}dy^idy^j=e^{2\phi}\hat{g}_{ab}dx^adx^b+G_{ij}dy^idy^j \,,\ee
where
\be G_{ab}=G_{ab}(x)\,,\quad G_{ij}=G_{ij}(x\,,y)\,,\quad G_{ai}=G^{ai}=0 \,.\ee
We are interested in evaluating the curvature tensors at a given codimension$-2$ hypersurface $\Sigma$ along $y$ directions. One has the metric expansion
\be G_{ij}(x\,,y)=\gamma_{ij}(y)+2K_{aij}x^a+Q_{abij}x^a x^b+\cdots\,,\ee
where $K_{aij}$ are extrinsic curvatures of $\Sigma$. Notice that under the above coordinates, $K_{aij}=\fft 12 \partial_a G_{ij}$ and $Q_{abij}\equiv \partial_a K_{bij}$. We use the metric $\hat{g}^{ab}$ and $\hat{g}_{ab}$ to raise and lower indices $a\,,b\,,\cdots$ for curvature tensors, for example $K^a_{ij}=\hat{g}^{ab}K_{bij}$. The inverse metric $G^{ij}$ is given by
\be G^{ij}=\gamma^{ij}-2K_{a}^{\,\,\,\,ij}x^a-\big(Q_{ab}^{\quad ij}-4K_{a}^{\,\,\,\,i \ell}K_{b\ell}^{\quad j} \big)x^a x^b+\cdots\,, \ee
where $K_a^{ij}=-\fft12 \partial_a G^{ij}$, where the minus sign emerges owing to the requirement $G^{ik}G_{kj}=\delta^i_{\,\,\,j}$.
However, we are interested in deriving Riemann curvatures on the surface $\Sigma$. In many cases, we can take the approximation
$G^{ij}(x\,,y)=\gamma^{ij}(y)$ but it is not always true. We need keep in mind that all the relevant terms around the surface $\Sigma$ should be included when necessary. Without confusion, we will raise and lower the indices $i\,,j\,,\cdots$ by using the induced metric $\gamma_{ij}$ on the surface $\Sigma$.

The Christoffel connection and the Riemann tensor can be calculated from their standard definitions. One has
\bea
&& R_{abcd}=r_{abcd}\,,\quad R_{iabc}=0\,,\quad R_{aijk}=\mc{D}_k K_{aij}-\mc{D}_j K_{aik} \,,\nn\\
&& R_{ijab}=\gamma^{kl}(K_{bik}K_{aj\ell}-K_{aik}K_{bj\ell}) \,,\nn\\
&& R_{iajb}=\Gamma^c_{ab}K_{cij}+\gamma^{k\ell}K_{ajk}K_{bi\ell}-Q_{abij}\,,\nn\\
&& R_{ijk\ell}=\mc{R}_{ijk\ell}+e^{-2\phi}(K_{ai\ell}K^a_{jk}-K_{aik}K^a_{j\ell})\,,
\eea
where
\be \Gamma^a_{bc}=2\delta^a_{(b}\partial_{c)} \phi-\partial^a \phi\, \hat{g}_{bc} \,.\ee
The Ricci tensor and scalar can be derived as
\bea
&& R_{ai}=\mc{D}^j K_{aji}-\partial_i K_a\,,\nn\\
&& R_{ab}=r_{ab}+\Gamma^c_{ab}K_c+K_{aij}K_b^{ij}-\gamma^{ij}Q_{abij}\,,\nn\\
&& R_{ij}=\mc{R}_{ij}+e^{-2\phi}\big(2K_{a\ell i}K^{a\ell j}-K_a K^a_{ij}-Q_{a\,\,\,\,ij}^{\,\,\,\,a}  \big)\,,
\eea
and
\be R=r+\mc{R}-e^{-2\phi}\big(K_aK^a-3K_{aij}K^{aij}+2\gamma^{ij}Q_{a\,\,\,\,ij}^{\,\,\,\,a} \big) \,.\ee
Note that the above relation is nothing else but an equivalent expression for Gauss-Codazzi identity. The Riemann tensor in the cone directions up to linear order in $\epsilon$ is given by
\be R_{abcd}=r_{abcd}=e^{2\phi}\big(\hat{g}_{ad}\partial_b\partial_c \phi+\hat{g}_{bc}\partial_a\partial_d \phi-\hat{g}_{ac}\partial_b\partial_d \phi-\hat{g}_{bd}\partial_a\partial_c
 \phi\big)+O(\epsilon^2) \,,\ee
where $\phi=-\epsilon \log{\rho}$ (without introducing a cut-off away from the cone $\rho=0$).
 This implies that $r_{ab}=-\hat{g}_{ab}\hat{\nabla}^2 \phi$ and $r=-2e^{-2\phi}\hat{\nabla}^2\phi$, where $\hat{\nabla}^2$ is defined with respect to the two dimensional metric $\hat{g}$. For later convenience, the metric $\hat{g}$ will be expressed frequently in three types of coordinates: the Cartesian coordinates $(x^1\,,x^2)$, the cylindrical coordinates $(\rho\,,\tau)$ and the complex coordinates $(z\,,\bar z)$, which are defined as
\bea
&&x^1=\rho \cos\tau\,,\quad x^2=\rho\sin\tau\,,\nn\\
&&z=\rho e^{i\tau}=x^1+ix^2\,,\quad \bar z=\rho e^{-i\tau}=x^1-ix^2\,.
\eea
One has
\be d\hat{s}^2_{(2)}=\delta_{ab}dx^adx^b=d\rho^2+\rho^2d\tau^2=dz d\bar z\,.\ee
To exclude contributions from the conical singularity of the orbifold geometry $\hat{\mc{M}}_n$, the function $\phi$ should be properly regularized, for example we may set $\phi=-\ft{\epsilon}{2}\log{(\rho^2+a^2)}$ and take the limit $a\rightarrow 0$ in the final. For example, $\partial_z \phi=-\epsilon/2z\,,\partial_{\bar z}\phi=-\epsilon/2\bar z$ and $\partial_z\partial_{\bar{z}}\phi=\pi\,\epsilon\,\delta^2(x)$. These relations determine the singular terms in the Riemann curvatures
\bea\label{curvature1}
&& R_{z\bar{z}z\bar{z}}=e^{2\phi}\partial_z\partial_{\bar{z}}\phi+\cdots \,,\nn\\
&& R_{izjz}=2\partial_z \phi\, K_{zij}+\cdots \,,\nn\\
&& R_{i\bar{z}j\bar{z}}=2\partial_{\bar z} \phi\, K_{\bar{z}ij}+\cdots  \,,
\eea
as well as those in the Ricci tensors
\bea\label{curvature2}
&& R_{zz}=2\partial_z \phi K_z+\cdots\,,\nn\\
&& R_{\bar{z}\bar{z}}=2\partial_{\bar z} \phi K_{\bar z}+\cdots\,,\nn\\
&& R_{z\bar{z}}=r_{z\bar{z}}+\cdots=-\pi\epsilon\,\delta^2(x)+\cdots\,.
\eea
In the above, the first two equalities imply that the trace of the extrinsic curvatures should vanish for Einstein's gravity. This proves the RT formula. Finally, for Ricci scalar
\be\label{curvature3} R=r+\cdots=e^{-2\phi}\Big[-4\pi\epsilon\,\delta^2(x)\Big]+\cdots \,. \ee
These results are sufficient to derive holographic entanglement entropy for general Riemannian gravities \cite{Dong:2013qoa}.

As a simple application, we take $\Sigma$ as a bifurcate event horizon, which has vanishing extrinsic curvatures
$K_{aij}=0$. In this case, all the singular contributions to the curvature tensors come from the conical two dimensions. The results can be expressed compactly as
\bea
&& R_{\mu\lambda\nu\rho}=\bar{R}_{\mu\lambda\nu\rho}-2\pi\epsilon\,\varepsilon_{\mu\lambda}\varepsilon_{\nu\rho}\,\delta_\Sigma\,,\nn\\
&& R_{\mu\nu}=\bar{R}_{\mu\nu}-2\pi\epsilon\, n_{\mu\nu}\,\delta_\Sigma\,,\nn\\
&& R=\bar{R}-4\pi \epsilon\,\delta^2(x)\,,
\eea
where $\bar{R}$ stands for the curvatures in the smooth region of $\hat{\mc{M}}_n$; $n^\mu_i$ are two normal vectors of $\Sigma$ (here the normal vectors are normalized to unity). $\varepsilon_{\mu\nu}=n_\mu^1 n_\nu^2-n_\mu^2 n_\nu^1$ is binormal vector of $\Sigma$ and $n_{\mu\nu}=\Sigma_{i=1}^2 n_\mu^i n_\nu^i$. Some useful relations are $n_\mu n^\mu=2=n_{\mu\nu}n^{\mu\nu}$ and
\be \varepsilon_{\mu\lambda}\varepsilon_{\nu\rho}=n_{\mu\nu} n_{\lambda \rho}-n_{\mu \rho}n_{\nu \lambda}\,,\quad \varepsilon_{\mu\sigma}\varepsilon_{\nu}^{\,\,\,\sigma}=n_{\mu \nu} \,.\ee
In the $n\rightarrow 1$ limit, the normal vectors in the complex coordinates are given by
\bea
&&n_1=\ft{1}{2} (dz+d\bar z)\,,\nn\\
&&n_2=\ft{1}{2i} (dz-d\bar z)\,.
\eea
This leads to $\varepsilon_{z\bar{z}}=i /2$ and
\be n_{zz}=0=n_{\bar z\bar z}\,,\quad n_{z\bar z}=\ft{1}{2} \,.\ee
These results are particularly useful to transform the entropy formula for higher derivative gravities into covariant forms.

\section{Derivation details about $(\ref{a31})$}

We list our results for each term in (\ref{a3final}) as follows
\bea
&& R^3\,:\qquad E^{\mu\nu\rho\sigma}=3R^2 g^{\rho[\mu}g^{\nu]\sigma} \,,\nn\\
&& RR_{\mu\nu}^2\,: \qquad E^{\mu\nu\rho\sigma}=g^{\rho[\mu}g^{\nu]\sigma} R^2_{\alpha\beta}+R\big(g^{\rho[\mu}R^{\nu]\sigma} -g^{\sigma[\mu}R^{\nu]\rho}\big)\,,\nn\\
&& RR_{\mu\nu\rho\sigma}^2\,: \qquad E^{\mu\nu\rho\sigma}=2R R^{\mu\nu\rho\sigma}+g^{\rho[\mu}g^{\nu]\sigma} R^2_{\alpha\beta\gamma\delta}\,,\nn\\
&& R_{\mu\sigma}R_{\nu}^{\,\,\,\,\sigma}R^{\mu\nu}\,: \qquad E^{\mu\nu\rho\sigma}=\fft32\big(
g^{\rho[\mu}R^{\nu]}_{\,\,\,\,\alpha} R^{\alpha\sigma} -g^{\sigma[\mu}R^{\nu]}_{\,\,\,\,\alpha} R^{\alpha\rho}\big)\,,\nn\\
&&R_{\mu\rho}R_{\nu\sigma} R^{\mu\nu\rho\sigma}\,: \qquad E^{\mu\nu\rho\sigma}=
R^{\rho[\mu}R^{\nu]\sigma}+R_{\alpha\beta}\big(g^{\rho[\mu}R^{\nu]\alpha\sigma\beta} -g^{\sigma[\mu}R^{\nu]\alpha\rho\beta}\big)\,,\nn\\
&&R_{\mu\lambda}R^{\mu}_{\,\,\,\,\nu\rho\sigma} R^{\lambda\nu\rho\sigma} \,: \qquad E^{\mu\nu\rho\sigma}=-2R^{[\mu}_{\,\,\,\,\alpha}R^{\nu]\alpha\rho\sigma}+\fft12\big(
g^{\rho[\mu}R^{\nu]\alpha\beta\gamma} R^{\sigma}_{\,\,\,\,\alpha\beta\gamma} -g^{\sigma[\mu}R^{\nu]\alpha\beta\gamma} R^{\rho}_{\,\,\,\,\alpha\beta\gamma}\big)\,,\nn\\
&& R_{\mu\nu}^{\quad\alpha\beta}R_{\alpha\beta}^{\quad\rho\sigma}R_{\rho\sigma}^{\quad \mu\nu}\,: \qquad E^{\mu\nu\rho\sigma}=3 R^{\mu\nu\alpha\beta}R_{\alpha\beta}^{\quad \rho\sigma}\,,\nn\\
&& R_{\mu\,\,\,\,\nu}^{\,\,\,\,\lambda\,\,\,\,\rho}
R_{\lambda\,\,\,\,\rho}^{\,\,\,\,\alpha\,\,\,\,\beta}R^{\,\,\,\,\mu\,\,\,\,\nu}_{\alpha\,\,\,\,\beta\,\,\,\,}\,: \qquad E^{\mu\nu\rho\sigma}=\fft32\big(R^{\alpha\rho\beta[\mu}R^{\nu]\,\,\,\,\sigma}_{\,\,\,\,\beta\,\,\,\,\alpha}
-R^{\alpha\sigma\beta[\mu}R^{\nu]\,\,\,\,\rho}_{\,\,\,\,\beta\,\,\,\,\alpha}  \big)\,.
\eea
It is straightforward to evaluate the Wald entropy terms
\bea
&& R^3\,:\qquad \mm{Wald}=-2\pi\int_\Sigma 6R^2 \,,\nn\\
&& RR_{\mu\nu}^2\,: \qquad \mm{Wald}=-2\pi\int_\Sigma 2\big( R_{\mu\nu}^2+R R_{\mu\nu}n^{\mu\nu} \big)\,,\nn\\
&& RR_{\mu\nu\rho\sigma}^2\,: \qquad \mm{Wald}=-2\pi\int_\Sigma 2\big( R_{\mu\nu\rho\sigma}^2+R R_{\mu\nu\rho\sigma}\varepsilon^{\mu\nu}\varepsilon^{\rho\sigma} \big)\,,\nn\\
&& R_{\mu\sigma}R_{\nu}^{\,\,\,\,\sigma}R^{\mu\nu}\,: \qquad \mm{Wald}=-2\pi\int_\Sigma 3R_{\mu\sigma}R_{\nu}^{\,\,\,\,\sigma}n^{\mu\nu}\,,\nn\\
&&R_{\mu\rho}R_{\nu\sigma} R^{\mu\nu\rho\sigma}\,: \qquad \mm{Wald}=-2\pi\int_\Sigma \big(R_{\mu\rho}R_{\nu\sigma}\varepsilon^{\mu\nu}\varepsilon^{\rho\sigma}+2R_{\mu\lambda\nu\rho}R^{\mu\nu}n^{\lambda\rho} \big)\,,\nn\\
&&R_{\mu\lambda}R^{\mu}_{\,\,\,\,\nu\rho\sigma} R^{\lambda\nu\rho\sigma} \,: \qquad \mm{Wald}=-2\pi\int_\Sigma \big( R_{\mu\alpha\beta\gamma}R_{\nu}^{\,\,\,\,\alpha\beta\gamma}n^{\mu\nu}+2R_{\mu}^{\,\,\,\,\lambda}R_{\lambda\nu\rho\sigma}\varepsilon^{\mu\nu}\varepsilon^{\rho\sigma} \big)\,,\nn\\
&& R_{\mu\nu}^{\quad\alpha\beta}R_{\alpha\beta}^{\quad\rho\sigma}R_{\rho\sigma}^{\quad \mu\nu}\,: \qquad \mm{Wald}=-2\pi\int_\Sigma 3R_{\mu\nu}^{\quad\alpha\beta}R_{\alpha\beta\rho\sigma}\varepsilon^{\mu\nu}\varepsilon^{\rho\sigma}\,,\nn\\
&& R_{\mu\,\,\,\,\nu}^{\,\,\,\,\lambda\,\,\,\,\rho}
R_{\lambda\,\,\,\,\rho}^{\,\,\,\,\alpha\,\,\,\,\beta}R^{\,\,\,\,\mu\,\,\,\,\nu}_{\alpha\,\,\,\,\beta\,\,\,\,}\,: \qquad \mm{Wald}=-2\pi\int_\Sigma 3R_{\mu\alpha\rho\beta}R_{\nu\,\,\,\,\sigma}^{\,\,\,\,\alpha\,\,\,\,\beta}\varepsilon^{\mu\nu}\varepsilon^{\rho\sigma}\,.
\eea
For RT surfaces in AdS vacuum, the results can be even more simplified. We shall not list them here.

For the anomaly terms, after straightforward but lengthy derivations, we obtain
 \bea
&& R^3\,:\qquad \ft{\partial^2 L}{\partial R_{izjz} \partial R_{k\bar{z}l \bar{z}}}=0 \quad \Longrightarrow \quad  \mm{Anomaly}=0\,,\nn\\
&& RR_{\mu\nu}^2\,:\qquad \ft{\partial^2 L}{\partial R_{izjz} \partial R_{k\bar{z}l \bar{z}}}=\fft12g^{ij}g^{k\ell}R \quad \Longrightarrow \quad \mm{Anomaly}=\fft{2\pi}{5} \int_\Sigma\, R\, K_aK^a \,,\nn\\
&& RR_{\mu\nu\rho\sigma}^2\,: \qquad \ft{\partial^2 L}{\partial R_{izjz} \partial R_{k\bar{z}l \bar{z}}}=2g^{ik}g^{j\ell}R \quad \Longrightarrow \quad \mm{Anomaly}=\fft{8\pi}{5} \int_\Sigma\, R\, K_{aij}K^{aij}   \,,\nn\\
&& R_{\mu\sigma}R_{\nu}^{\,\,\,\,\sigma}R^{\mu\nu}\,: \qquad  \ft{\partial^2 L}{\partial R_{izjz} \partial R_{k\bar{z}l \bar{z}}}=\fft34g^{ij}g^{k\ell}R^a_{\,\,\,\,a} \quad \Longrightarrow \quad
\mm{Anomaly}=\fft{3\pi}{2} \int_\Sigma\, R^a_{\,\,\,\,a} K_bK^b   \,,\nn\\
&&\nn\\
&&R_{\mu\rho}R_{\nu\sigma} R^{\mu\nu\rho\sigma}\,: \qquad \ft{\partial^2 L}{\partial R_{izjz} \partial R_{k\bar{z}l \bar{z}}}=\fft12\big(g^{ij}R^{k\ell}+g^{k\ell}R^{ij}+4g^{ij}g^{k\ell}R_{z\bar{z}z\bar{z}} \big)\nn\\
&&\qquad\qquad\qquad\qquad \quad  \mm{Anomaly}=2\pi\int_{\Sigma}\Big(\fft23 K^aK_{aij}R^{ij}-\fft12 R^{ab}_{\quad ab}\,K_cK^c\Big)   \,,\\
&&\nn\\
&&R_{\mu\lambda}R^{\mu}_{\,\,\,\,\nu\rho\sigma} R^{\lambda\nu\rho\sigma} \,: \qquad \ft{\partial^2 L}{\partial R_{izjz} \partial R_{k\bar{z}l \bar{z}}}=\fft12\big(g^{ij}R^{kz\ell\bar z}+g^{k\ell}R^{i\bar{z}jz}+g^{ik}g^{j\ell}R^{z\bar z} \big)+g^{j\ell}R^{ik} \nn\\
&&\qquad\qquad\qquad\qquad\qquad  \mm{Anomaly}=2\pi\int_\Sigma\,\Big(\fft12 K_{aij}K^{aij}R^b_{\,\,\,\,b}+ K^aK_{aij}R^{i\,\,\,\,jb}_{\,\,\,\,b}+\fft23 K_{a\ell i}K^{a\ell j}R_{j}^{\,\,\,\,i}\Big)  \,,\nn\\
&&\nn\\
&& R_{\mu\nu}^{\quad\alpha\beta}R_{\alpha\beta}^{\quad\rho\sigma}R_{\rho\sigma}^{\quad \mu\nu}\,: \qquad \ft{\partial^2 L}{\partial R_{izjz} \partial R_{k\bar{z}l \bar{z}}}=3\big(g^{ik}R^{jz\ell\bar z} +g^{j\ell}R^{izk\bar z}\big) \nn\\
&&\qquad\qquad\qquad\qquad\qquad \quad   \mm{Anomaly}=2\pi\int_\Sigma 6 K_{ai\ell}K_{b\,\,\,\,j}^{\,\,\,\,\ell}R^{\,\,\,\,i\,\,\,\,j}_{c\,\,\,\,d}\big(n^{ab}n^{cd}+\varepsilon^{ab}\varepsilon^{cd} \big)\,,\nn\\
&&\nn\\
&& R_{\mu\,\,\,\,\nu}^{\,\,\,\,\lambda\,\,\,\,\rho}
R_{\lambda\,\,\,\,\rho}^{\,\,\,\,\alpha\,\,\,\,\beta}R^{\,\,\,\,\mu\,\,\,\,\nu}_{\alpha\,\,\,\,\beta\,\,\,\,}\,: \qquad
\ft{\partial^2 L}{\partial R_{izjz} \partial R_{k\bar{z}l \bar{z}}}=\fft32R^{ik j\ell} + \fft38 g^{ik}g^{j\ell}R^{z\bar{z}z\bar{z}}+
\fft34\big(g^{ik}R^{j\ell z\bar{z}}+g^{j\ell}R^{ik z\bar{z}} \big) \nn\\
&&\qquad\qquad\qquad\qquad\qquad\quad  \mm{Anomaly}=2\pi\int_\Sigma \Big(-\fft32 K_{aij}K^{aij}R^{cd}_{\quad cd}+\fft35 K_{aij}K^{ak\ell}R^{i\,\,\,\,j}_{\,\,\,\,k\,\,\,\,\ell}+2K_{a\ell i}K_{b\,\,\,\,j}^{\,\,\,\,\ell}R^{ijab}\Big)    \,.\nn
\eea
 The results can be greatly simplified by using the fact that $\Sigma$ is extremal and hence $K_a=0$. Moreover, the spacetime is pure AdS so that $R^a_{\,\,\,\,a}=-2(D-1)\ell^{-2}_{AdS}\,,R^{ab}_{\quad ab}=-2\ell^{-2}_{AdS}$. It follows that
\bea
&& R^3\,,RR_{\mu\nu}^2\,,R_{\mu\sigma}R_{\nu}^{\,\,\,\,\sigma}R^{\mu\nu}\,,R_{\mu\rho}R_{\nu\sigma} R^{\mu\nu\rho\sigma}\,:\quad \mm{Anomaly}=0\,,\nn\\
&& RR_{\mu\nu\rho\sigma}^2\,: \quad  \mm{Anomaly}= -\ft{8\pi}{5}\ell^{-2}_{AdS} \int_\Sigma\, D(D-1) K_{aij}K^{aij}  \,,\nn\\
&&R_{\mu\lambda}R^{\mu}_{\,\,\,\,\nu\rho\sigma} R^{\lambda\nu\rho\sigma} \,: \quad\mm{Anomaly}=-\ft{10\pi}{3}\ell^{-2}_{AdS} \int_\Sigma\, (D-1) K_{aij}K^{aij}  \,,\nn\\
&&\nn\\
&& R_{\mu\nu}^{\quad\alpha\beta}R_{\alpha\beta}^{\quad\rho\sigma}R_{\rho\sigma}^{\quad \mu\nu}\,: \quad  \mm{Anomaly}=-24\pi \ell^{-2}_{AdS} \int_\Sigma K_{aij}K^{aij}\,,\nn\\
&&\nn\\
&& R_{\mu\,\,\,\,\nu}^{\,\,\,\,\lambda\,\,\,\,\rho}
R_{\lambda\,\,\,\,\rho}^{\,\,\,\,\alpha\,\,\,\,\beta}R^{\,\,\,\,\mu\,\,\,\,\nu}_{\alpha\,\,\,\,\beta\,\,\,\,}\,:
  \quad \mm{Anomaly}=\ft{36\pi}{5}\ell^{-2}_{AdS} \int_\Sigma\,  K_{aij}K^{aij}    \,.\nn
\eea
Combing all the results above, one finally arrives at $(\ref{a31})$.

\end{document}